\documentclass[%
 reprint,
 amsmath,amssymb,
 aps,
 pra,
]{revtex4-2}

\usepackage{graphicx}
\usepackage{subfig}
\usepackage{dcolumn}
\usepackage{bm}
\usepackage{braket}
\usepackage{color}
\usepackage{siunitx}
\usepackage{xfrac}
\usepackage{tabularx}
\usepackage{hyperref}

\begin{document}

\preprint{APS/123-QED}

\title{Generation of photon pairs by spontaneous four-wave mixing in linearly uncoupled resonators}

\author{Luca Zatti}
 \affiliation{Department of Physics, University of Pavia, I-27100 Pavia, Italy}
 \author{J.E. Sipe}
 \affiliation{Department of Physics, University of Toronto, 60 St. George St., Toronto, Ontario M5S 1A7, Canada}%
\author{Marco Liscidini}%
\affiliation{Department of Physics, University of Pavia, I-27100 Pavia, Italy}%

\begin{abstract}
We present a detailed study of the generation of photon pairs by spontaneous four-wave mixing in a structure composed of two linearly uncoupled resonators, where energy can be transferred from one resonator to another only through a nonlinear interaction. Specifically, we consider the case of two racetrack-shaped resonators connected by a coupler designed to guarantee that the resonance comb of each resonator can be tuned independently, and to allow the nonlinear interaction between modes that belong to different combs. We show that such a coupler can be realized in at least two ways: a directional coupler or a Mach-Zehnder interferometer. For these two scenarios, we derive analytic expressions for the pair generation rate via single-pump spontaneous four-wave mixing, and compare these results with that achievable in a single ring resonator.  
\end{abstract}

\maketitle

\section{Introduction}

Many sources of nonclassical light are based on parametric fluorescence processes, such as spontaneous parametric-down conversion (SPDC) and spontaneous four-wave mixing (SFWM). 
Initially, SPDC and SFWM were studied in bulk systems, such as crystals \cite{Harris:67,Burnham1970} and optical fibers \cite{Inoue2004,Li:04}, with the use of intense pump fields. Yet their efficiency can be increased by several orders of magnitudes with the use of nanostructures,  which leverage the electromagnetic field enhancement associated with the spatial and temporal light confinement in resonant structures \cite{clemmen2009continuous,caspani2017integrated}. Indeed, in the last fifteen years the generation of photon pairs by SPDC or SFWM has been demonstrated in photonic devices employing several materials, including silicon \cite{azzini2012classical} , silicon nitride \cite{Imany:18}, Hydex \cite{kues2017chip}, and III-V semiconductors \cite{ducci2013}. All these platforms are currently being investigated, with the aim of developing fully integrated photonic systems for applications in several areas of quantum technologies, such as computation, simulation, and communication.

While the first efforts were mainly focused on improving the pair generation rate,  subsequent studies have also investigated the use of design strategies to engineer the properties of the generated pairs by optimizing the structure of the integrated photonic device.
These studies range from the application of waveguide dispersion engineering to achieve phase matching in a desired frequency range \cite{sharping2006generation,takesue2007entanglement}, to the use of more complex systems composed of two or more coupled resonators to form photonic molecules \cite{Gentry2014,Zeng2015}, and to the use of lattices \cite{davanco2012telecommunications,xiong2011slow} that can exhibit topological properties \cite{Mittal2018,Blanco2018}. The overall strategy is to engineer the electromagnetic field enhancement to favor the generation of photons in particular modes, and to inhibit it in others. Finally, structure design can be useful in suppressing unwanted parasitic nonlinear processes that can be sources of noise \cite{zhang2021}.

A few years ago, Menotti $et al.$ showed that parametric nonlinear interactions can occur in systems composed of two or more integrated resonators that are linearly uncoupled \cite{menotti2019nonlinear}. In this kind of structure, each normal mode can be associated with one resonator, and energy passes from one mode to the other only thanks to the presence of a nonlinear interaction. This can occur because two or more normal modes that are nonlinearly coupled overlap in part of the structure. On the one hand, the strength of the nonlinear interaction is reduced compared to what it would be were the modes sharing the full structure, 
e.g., for SFWM involving the modes of a single ring resonator. On the other hand, having the resonators linearly uncoupled makes it easier to engineer their spectral properties and resonant field enhancement. This is particularly useful for nonlinear optical processes, which typically require several conditions to be met, from those necessary to guarantee the process efficiency (e.g. phase-matching) to those needed to suppress unwanted parasitic processes. To date, the experimental realization of this approach has made use of racetrack resonators placed side-by-side forming a directional coupler (DC), with  stimulated and spontaneous four-wave mixing demonstrated in silicon nitride \cite{Tan:20} and silicon \cite{Preble20,sabattoli2021} resonators, respectively. 


In this work we study two systems. In one the coupling between the two racetrack resonators is provided by a DC, as mentioned above; in the other it is provided by a Mach-Zehnder interferometer (MZI). Each is designed to ensure the linear uncoupling of the two resonators, and yet to maximize their nonlinear interaction for single-pump SFWM. In Sec. II we present the structures and analyze their linear properties. In Sec. III we evaluate the pair generation rates, taking into account scattering losses in the structures. In Sec. IV we compare the SFWM efficiencies of the different structures with that of a single ring resonator. Finally, in Sec V we draw our conclusions. Calculational details are presented in four appendices, in the first of which we also give the expression for the biphoton wave function of the pairs that survive scattering losses and exit the structures.

\section{\label{sec:level1}Structure and linear properties\protect}

\begin{figure}[t]
\includegraphics[width=0.45\textwidth]{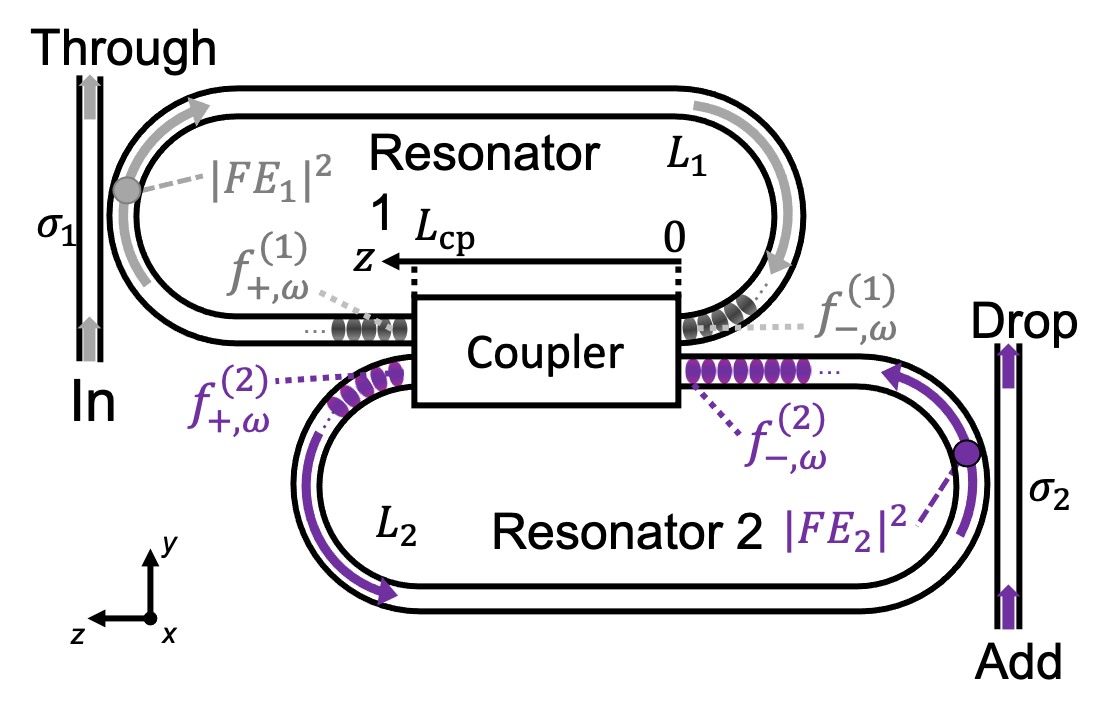}
\caption{\label{fig:Structure} A sketch of the double racetrack resonator structure.}
\end{figure}

We begin by considering structures of the general form sketched in Fig. \ref{fig:Structure}, where there are two racetrack resonators and a coupling region, the ``coupler," between them. Each resonator is also point-coupled to a bus waveguide, where $\sigma_{1,m(2,m)}$ is the waveguide self-coupling coefficient between the bus waveguide and the resonator 1 (2)  at frequencies in the neighborhood of its $m^{th}$ resonance, with  $0\leq\sigma_{i,m}\le1$  \cite{photYY,heebner2008optical}. 
We assume the bus waveguides and the waveguides of the resonators to be the same, and single mode at the frequencies of interest. We characterize them by a complex propagation constant $\tilde{k}(\omega)=k(\omega) +i\xi/2$, where $\xi$ is a frequency independent phenomenological constant that takes into account the scattering losses associated with light propagation \cite{banic2022two}, and
\begin{align}
k(\omega) = k_0 + \frac{1}{v_g}(\omega-\omega_0)+\frac{1}{2}\beta_2 (\omega-\omega_0)^2 + \cdots,
\end{align}
is the real part of the propagation constant expanded in Taylor series around a reference frequency $\omega_0$. Here $k_0 = n_{\text{eff}} \, \omega_0/c$, with $n_{\text{eff}}$ being the effective index at $\omega_0$, $v_g = [d\omega /dk]_{\omega_0}$ being the group velocity, and $\beta_2 = [d^2k/d\omega^2]_{\omega_0}$ being the group velocity dispersion (GVD). In the following we will consider the expansion up to the first order, neglecting the GVD and the higher terms over the full range of frequencies considered.
We are interested in couplers that can be designed so that in the linear regime there is \textit{no} coupling. That is, in this regime, light entering the coupler from one resonator exits into the \textit{same} resonator. The nonlinear interaction between these modes is discussed in Sec. III.

Very generally, in the linear regime a coupler linking the two resonators
can be described by a unitary matrix $X$ that links the input fields $f_{-,\omega}^{(1)}$ and $f_{-,\omega}^{(2)}$ at the beginning of the coupler ($z=0$) to the output fields $f_{+,\omega}^{(1)}$ and $f_{+,\omega}^{(2)}$ at the end of it ($z = L_{\text{cp}}$) as
\begin{align}\label{UnitaryMatrix}
    \begin{pmatrix}
     f_{+,\omega}^{(1)}\\
     f_{+,\omega}^{(2)}
    \end{pmatrix}
    = X
    \begin{pmatrix}
     f_{-,\omega}^{(1)} \\
     f_{-,\omega}^{(2)}
    \end{pmatrix} 
    = 
    \begin{pmatrix}
     X_{11} & X_{12} \\
     X_{21} & X_{22}
    \end{pmatrix}
    \begin{pmatrix}
     f_{-,\omega}^{(1)} \\
     f_{-,\omega}^{(2)}
    \end{pmatrix},
\end{align}
where $f_{\pm,\omega}^{(i)}$ is the field circulating in the $i^{th}$ resonator at angular frequency $\omega$. This matrix satisfies the relation $XX^{\dagger}=X^{\dagger}X = I$, with $\text{det}(X)=1$.
If we want the two resonators to be uncoupled, in a realistic situation the terms $X_{12}$ and $X_{21}$ should be as close as possible to zero.
Naturally, one could achieve high linear isolation of the two racetracks by designing a DC with very distant waveguides, but this would also prevent any nonlinear interaction between them. Instead, one can design the coupler region such that the modes of the two resonators share a spatial region inside the coupler, and yet the two resonators are uncoupled in the linear regime.

\begin{figure}[!t]
\includegraphics[width=0.48\textwidth]{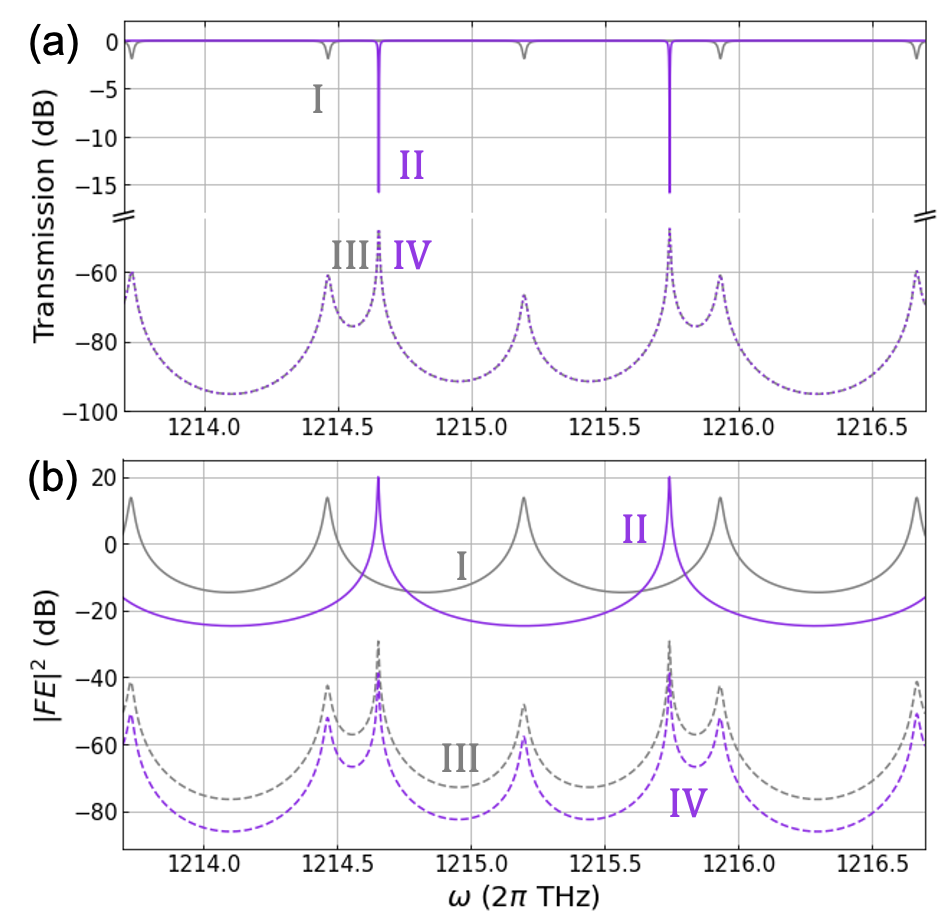}
\caption[justification=justified]{(a) Transmission spectra for In to Through (I), Add to Drop (II), In to Drop (III), and Add to Through (IV). (b) Corresponding intensity enhancement.
The dips in lines I and II in (a), which occur at the resonance frequencies, correspond to the intensity enhancement peaks of lines I and II shown in (b). We assumed $L_{1} = \SI{641}{\mu m}$, $L_{2} = \SI{432}{\mu m}$, $\sigma_{1} = 0.933$, $\sigma_{2} = 0.993$, $\xi = \SI{0.23}{cm^{-1}}$ (corresponding to $\SI{1}{dB/cm}$), and a value of the coupling coefficient $ X_{12} = X_{21} = -i0.00161$, the last value chosen to ensure that the resonators are nearly linearly uncoupled.} 
\label{fig:LinProp}
\end{figure}

In such a situation, each resonator has a well-defined set of resonances that are associated with light confinement mainly in it.
Each resonance frequency $\omega_{i,m}$ satisfies the usual condition $k(\omega_{i,m})L_i=2m\pi$, where we indicated with $L_{i}$ the lengths of the $i^{th}$ resonator and with $\omega_{i,m}$ the $m^{th}$ resonant frequency of the $i^{th}$ resonator. 
To characterize this structure in the linear regime, in Fig. \ref{fig:LinProp} we plot the transmission and the intensity enhancement, i.e., the field enhancement (FE) modulus squared, as a function of frequency for various in-out port configurations (see Fig. \ref{fig:Structure}). We assume a realistic case in which the two resonators are just \textit{nearly} linearly uncoupled, and consider a frequency range $|\omega-\omega_{i,m}|\ll{v_g}/{L_i}$. Then we can easily identify modes associated primarily with one or the other of the two resonators, and find more than \SI{30}{dB} on-resonance isolation between the two resonators. In a practical realization of such devices, a striking
advantage of this configuration is that one can control the relative position of the two resonance combs by means of electric heaters \cite{Tan:20}, or any other mechanism that induces an effective refractive index change in one of the resonators in a region far from the coupler. 

In the limit where the ``coupler" in fact provides no coupling between the resonators in the linear regime, if low intensity light is injected into the $i^{th}$ resonator through the corresponding bus waveguide, then at frequencies close to that of the $m^{th}$ resonance the intensity enhancement of the light injected from the bus waveguide into the resonator \cite{banic2022two} can be written as 
\begin{align}\label{ZingFE}
|\text{FE}_{i,m}(\omega)|^2 = |\text{FE}_{(\text{max})i,m}|^2 \frac{\frac{\Gamma_{i,m}^2}{4}}{\frac{\Gamma_{i,m}^2}{4}+(\omega-\omega_{i,m})^2},
\end{align}
where 
\begin{equation}\label{FEmax}
|\text{FE}_{(\text{max})i,m}|^2 = \frac{1-\sigma_{i,m}^2}{(1-\sigma_{i,m} a_{i,m})^2}
\end{equation}
is the maximum value of the intensity enhancement, and
\begin{equation}\label{Gamma}
\Gamma_{i,m} =\frac{v_g}{L_i}\frac{2(1-\sigma_{i,m} a_{i,m})}{\sqrt{\sigma_{i,m} a_{i,m}}}
\end{equation}
is the line width; here $1-a_i^2$ is the round trip loss of the $i^{th}$ resonator, where $a_i=e^{- \xi L_i}$. In this limit, at the critical coupling condition (i.e., $\sigma_{i,m}=a_{i,m}$), one has $|\text{FE}_{(\text{max})i,m}|^2\simeq\mathcal{F}_{i,m}/\pi$, while for 
overcoupling (i.e., $a_{i,m}\ll\sigma_{i,m}$), one has $|\text{FE}_{(\text{max})i,m}|^2\simeq 2\mathcal{F}_{i,m}/{\pi}$, where $\mathcal{F}_{i,m}=2\pi \text{FSR}_i/\Gamma_{i,m}$ is the so-called resonator \emph{finesse}, and $\text{FSR}_i=v_g/L_i$ is the free spectral range of the $i^{th}$ resonator.

Note that the field distribution inside the coupler is not relevant for the establishment of linear uncoupling, as long as it is guaranteed that light entering from one resonator is redirected into the same one.
However, if one is interested in achieving nonlinear coupling between the two resonators, the field distribution inside the coupling region is crucial, as the strength of the nonlinear interaction depends on the spatial integral of the involved fields, which can be nonvanishing only in the coupling region. 
The coupler can be realized in several ways, such as the use of a multi-mode interference coupler, and alternative schemes for implementing different nonlinear optical processes in linearly uncoupled resonators can be considered. In this work we study two possibilities for the structure: a DC and a MZI. We consider the use of the resulting devices for implementing SFWM, where pump light is injected in the In port at a resonance frequency of Resonator 1, and signal and idler light is generated at resonance frequencies of Resonator 2 and exits through the Drop port.

\subsection{Directional coupler}

\textbf{\begin{figure}[!t]
\includegraphics[width=0.46\textwidth]{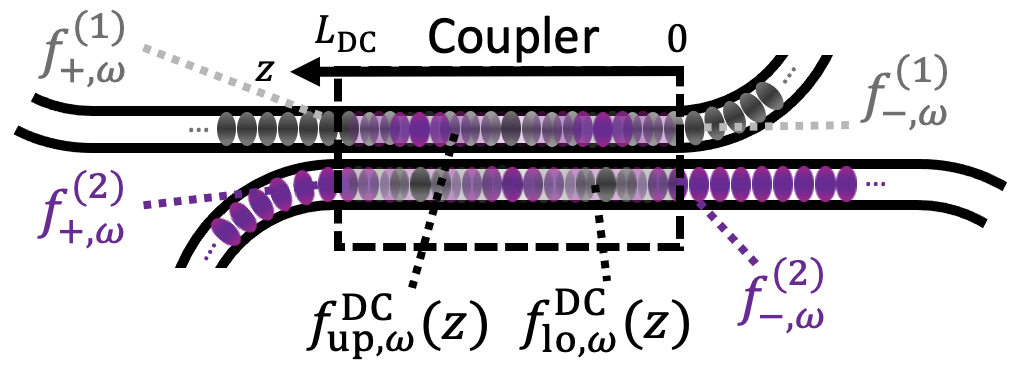}
\caption{\label{fig:DC} Sketch of the DC and schematic representation of the field distribution inside the channels.}
\end{figure}}

We first consider the structure where the coupler is a DC with length $L_{\text{DC}}$ (Fig. \ref{fig:DC}). The linear coupling between the two waveguides forming the DC can be described in the framework of standard coupled mode theory \cite{photYY}, in which the coupling constant $\kappa_{\text{DC}}$ depends on the linear overlap integral of the transverse field profile of the waveguide modes, which is a function of distance along the coupling region.
We restrict ourselves to a frequency range that is sufficiently small that $\kappa_{\text{DC}}$ can be considered frequency independent. Then when
$L_{\text{DC}}=n\pi/\kappa_{\text{DC}}$, with $n$ being a positive integer, 
the DC cross transmission is zero, yielding a high isolation of the two resonators in the linear regime \cite{menotti2019nonlinear}; in the absence of coupling to the bus waveguides, the energy of the resonant modes of the structure would be mainly confined to one resonator or the other.

We now take a closer look at the field distribution inside the DC, as sketched in Fig. \ref{fig:DC}. One can write the displacement field associated with each channel as
\begin{equation} \label{Displacement}
\mathbf{D}_{ch,\omega}(\mathbf{r}) = f_{ch,\omega}(z)\mathbf{d}_{ch}(x,y)\frac{e^{ik(\omega)z}}{\sqrt{2\pi}} \quad,
\end{equation}
where $ch=\text{up, lo}$, with ``up'' (``lo'') referring to the channel belonging to Resonator 1(2) as shown in Fig. \ref{fig:Structure}, and $\mathbf{d}_{ch}(x,y)$ is the displacement field distribution in the plane transverse to the propagation direction, properly normalized 
\cite{yang2008spontaneous}. As we take all the waveguides involved in the structure to be the same, we can assume that $\mathbf{d}_{ch}(x,y)$ is the same for all the channels under consideration; we also assume that it can be approximated as being independent of $\omega$. Finally, ${f}_{ch,\omega}(z)$ is a slowly varying envelope function that takes into account for the field variation along $z$; this function does not depend on the intensity of the light circulating in the structure but rather on the geometry of the coupler.
We have
\textbf{\begin{figure}[t]
\includegraphics[width=0.46\textwidth]{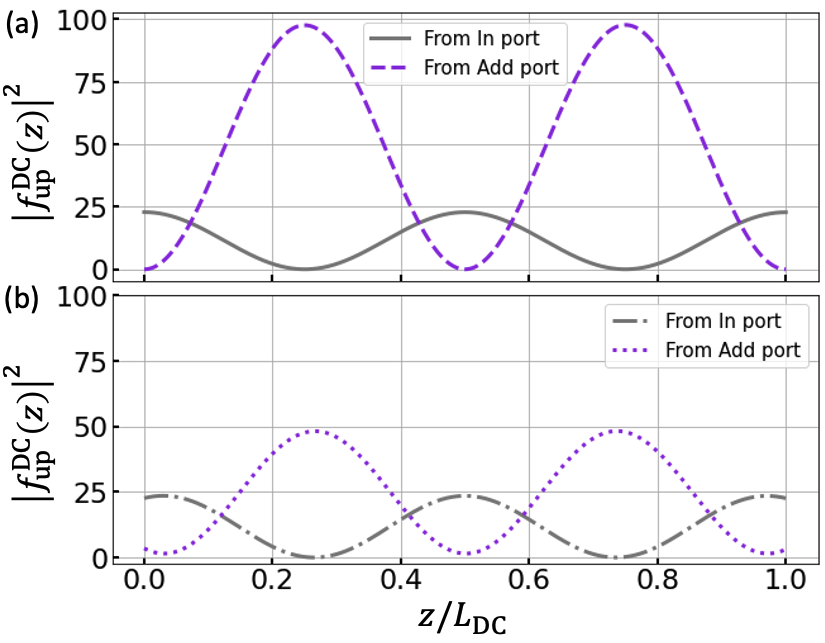}
\caption{\label{fig:Fields} Intensity distribution in the upper channel of the DC for two different values of the coupling coefficient: (a) $\kappa_\text{DC} = 0.064 \, \mu \text{m}^{-1}$ (perfect uncoupling) and (b) $\kappa_\text{DC} = 0.068 \, \mu \text{m}^{-1}$ (residual coupling).}
\end{figure}}
\begin{align}\label{DC-Ampl}
\begin{cases}
 {f}_{\text{up},\omega}^{\text{DC}}(z)  = \\ 
 \quad \quad {f}_{\text{up},\omega}(0) \cos(|\kappa_\text{DC}|z) - i{f}_{\text{lo},\omega}(0) \sin(|\kappa_\text{DC}|z) \\
 {f}_{\text{lo},\omega}^{\text{DC}}(z) = \\ 
 \quad \quad - i {f}_{\text{up},\omega}(0) \sin(|\kappa_\text{DC}|z) + {f}_{\text{lo},\omega}(0) \cos(|\kappa_\text{DC}|z) 
\end{cases}
,
\end{align}
with ${f}_{\text{up(lo)},\omega}(0)$ determined by the appropriate boundary conditions: ${f}_{\text{up(lo)},\omega}(0)$ = $f^\text{(1,2)}_{-,\omega} $, 
and ${f}_{\text{up(lo)},\omega}(L_\text{DC})$ = $f^\text{(1,2)}_{+,\omega} $. Then the coefficients of the unitary matrix \eqref{UnitaryMatrix} are found to be
\begin{align}
    &X_{11} = \cos|\kappa_\text{DC}|L_{\text{DC}}) , \\ 
    &X_{12} = - i \sin(|\kappa_\text{DC}|L_{\text{DC}}) , \\ 
    &X_{21} = - i \sin(|\kappa_\text{DC}|L_{\text{DC}}) , \\ 
    &X_{22} = \cos(|\kappa_\text{DC}|L_{\text{DC}}),  
\end{align}
where we note the overall phase $e^{i k(\omega) L_\text{DC}}$ due to the field propagation in the DC is included in the fast-varying component of \eqref{Displacement}.

In Fig. \ref{fig:Fields} we show the field profile inside the DC for two different coupling configurations.
We first consider the field distribution $|f_{up}(z)|^2$ in the upper channel for perfect linear uncoupling [see Fig. \ref{fig:Fields}(a)]. We consider a typical value of $\kappa_{\text{DC}} = \SI{0.064}{\mu m^{-1}}$, and we take $L_{\text{DC}}=2\pi/\kappa_{\text{DC}}=\SI{98.2}{\mu m}$. Light is injected into the In port (solid grey line) on resonance with Resonator 1 at $\omega_\text{In} = 1215.20 \times 2\pi$ THz ($\lambda_\text{In} = \SI{1550.07}{nm}$) and into the Add port (dashed violet line) on resonance with Resonator 2 at $\omega_\text{Add} = 1214.67 \times 2\pi$ THz ($\lambda_\text{Add} = \SI{1550.75}{nm}$; see Fig. \ref{fig:LinProp}). As expected, in this situation, at the beginning ($z = 0$) and at the end ($z = L_\text{DC}$) of the DC, $|f_\text{up}^\text{In}(z)|^2$ is maximum, while $|f_\text{up}^\text{Add}(z)|^2$ is zero. 

We now consider a small deviation from this ideal; we take the same $L_{\text{DC}}=\SI{98.2}{\mu m}$, but a larger coupling constant $\kappa_\text{DC} = 0.068 \, \mu \text{m}^{-1}$. This would arise, for example, if the waveguides were slightly closer to each other than in the ideal situation.
We plot the corresponding intensity distribution in Fig. \ref{fig:Fields}(b). Unlike in the ideal situation, here the intensity of the light injected into the Add port is slightly different from zero at the end of the waveguide, and that of the light injected into the In port is not quite at the maximum there, indicating a small linear coupling between the two resonators. More surprisingly, while the field intensity for the light associated with the mode of Resonator 1 is essentially unchanged, that associated with the mode of Resonator 2 is half of that shown in Fig. \ref{fig:Fields}(a). Such a remarkable difference demonstrates that the presence of some coupling between two resonators, which occurs if the DC is not ideal, does not affect all the modes in the same way. 
If we look at Fig. \ref{fig:LinProp}, we notice that the resonance associated with Resonator 2 at $1214.67 \times 2\pi$ THz is close to a resonance of Resonator 1, and thus even a small variation of the DC cross-coupling coefficient, such as that considered here, can lead to a strong reduction of the field intensity in Resonator 2. In contrast, the resonance of Resonator 1 at $1215.20 \times 2\pi$ THz is spectrally far from other resonances, which minimizes the linear coupling to those other modes.

These results show that a DC can be used to achieve the spatial overlap of modes belonging to linearly uncoupled [Fig. \ref{fig:Fields}(a)] or nearly uncoupled [Fig. \ref{fig:Fields}(b)] resonators. The approach is conceptually very simple, and can be realized in compact structures. However, one can identify two potential problems with this implementation. The first is that the DC properties critically depend on the value of $\kappa_\text{DC}$, which can be considered frequency independent only in a limited bandwidth, typically only a few tens of nanometers at telecom wavelengths. The second is that the field distributions of the modes belonging to different resonators are in quadrature along the DC, as shown in Fig. \ref{fig:Fields}, and thus any nonlinear interaction between them is expected to be small. In the following we introduce a different structure to overcome these limitations.

\subsection{Mach-Zehnder interferometer}

We now consider the Mach-Zehnder interferometer coupler, sketched in Fig. \ref{fig:MZI}, which is composed of two waveguides that are connected by two point couplers (PCs). The PCs are characterized by self-coupling coefficients $\sigma_\textit{sx}$ and $\sigma_\textit{dx}$ and cross-coupling coefficients $\kappa_\textit{sx}$ and $\kappa_\textit{dx}$, respectively \cite{photYY,heebner2008optical}. We take the coefficients to be real and positive; then from energy conservation $\sigma^2_\textit{sx(dx)}+\kappa^2_\textit{sx(dx)}=1$, and the splitting ratios of the PCs are defined as $(100\sigma^2_\textit{sx(dx)}):(100\kappa^2_\textit{sx(dx)})$. It follows 
that this system can be described by the unitary matrix \eqref{UnitaryMatrix}, with

\begin{figure}[t]
\includegraphics[width=0.46\textwidth]{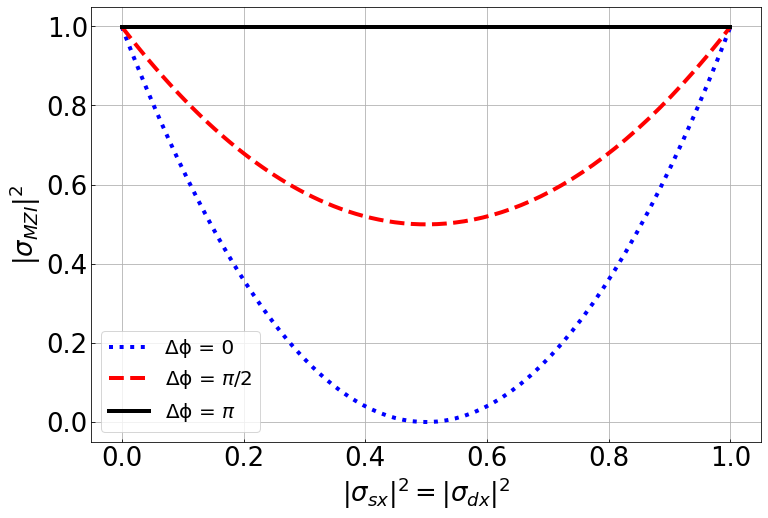}
\caption{\label{fig:sigmaMZ} Effective straight-through coefficient $\sigma_\text{MZI}$ of the interferometer for different phases $\Delta \phi$ as a function of the coupling coefficients with the two PCs assumed to be identical.}
\end{figure}

\textbf{\begin{figure}[t]
\includegraphics[width=0.46\textwidth]{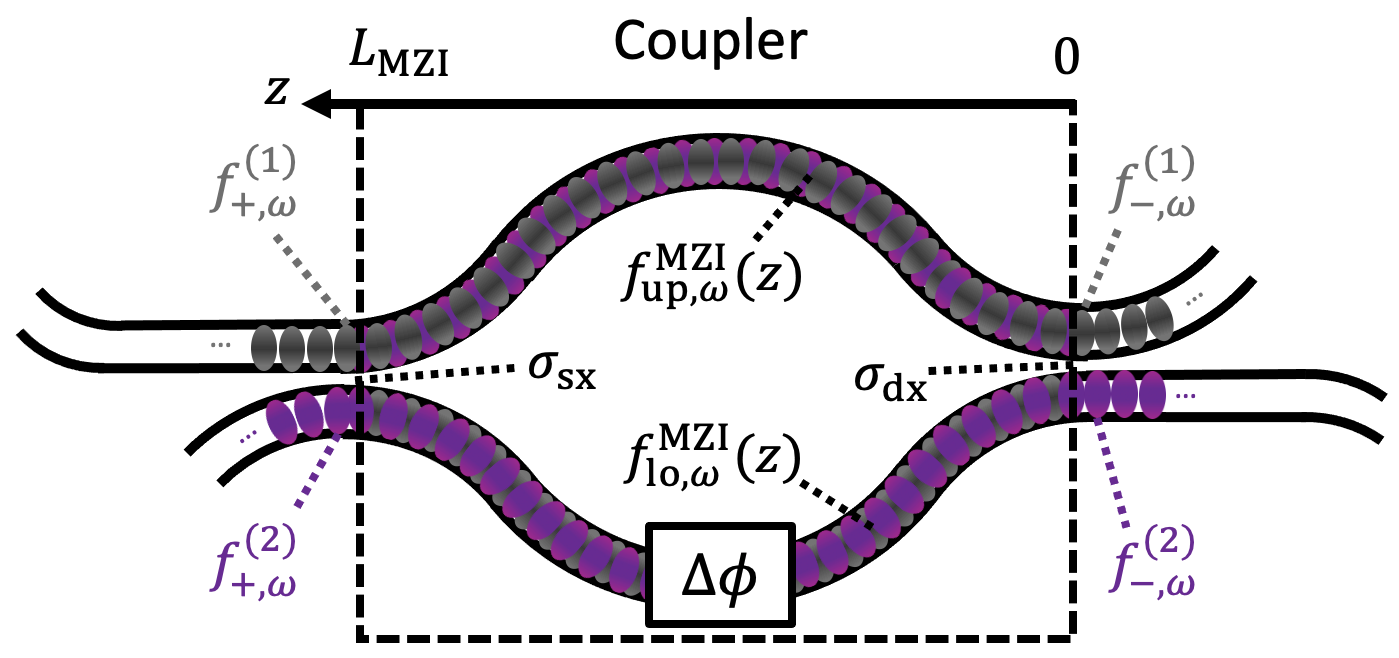}
\caption{\label{fig:MZI} Sketch of the Mach-Zehnder interferometer and a schematic representation of the overlap of the fields in the channels.}
\end{figure}}

\begin{align}
    &X_{11} = \sigma_{sx} \sigma_{dx}- \kappa_{sx}\kappa_{dx}e^{i\Delta \phi} , \\ 
    &X_{12} = i\left[ \sigma_{sx}\kappa_{dx} + \kappa_{sx}\sigma_{dx}e^{i\Delta \phi} \right] , \\ 
    &X_{21} = i\left[\kappa_{sx}\sigma_{dx} + \sigma_{sx}\kappa_{dx}e^{i\Delta \phi} \right] , \\ 
    &X_{22} = - \kappa_{sx}\kappa_{dx} + \sigma_{sx} \sigma_{dx}e^{i\Delta \phi},
\end{align}
where $\Delta \phi$ is the optical phase difference between the two arms of the interferometer. Thus the interferometer indeed acts as coupler, characterized by an effective straight-through coefficient 
\begin{align}\label{sigmaMZI}
\sigma_\text{MZI} = X_{11} & =   \sigma_\text{sx} \sigma_\text{dx} -\kappa_\text{sx} \kappa_\text{dx}e^{i \Delta \phi} \ ,
\end{align}
which identifies the fraction of field amplitude in Resonator 1 that is transferred back into it.
In Fig. \ref{fig:sigmaMZ} we show the modulus squared of $\sigma_\text{MZI}$, in the special case of $\sigma_{dx}=\sigma_{sx}=\sigma$, for different values of $\Delta\phi$. Interestingly, when $\Delta\phi=(2m+1)\pi$ (solid black line), the coefficient $\sigma_\text{MZI} = 1$ for any  value of $\sigma$. So
one can exploit interference at the output of the interferometer to achieve linear uncoupling of the two resonators. The curve shows that for a symmetric interferometer this is very robust: perfect linear uncoupling is obtained as long as the PCs are identical.

We now turn to the field inside each arm of the interferometer, again described by Eq. \eqref{Displacement}. Here the slowly varying envelope function ${f}_{\text{up}(\text{lo}),\omega}^\text{MZI}(z)$ in the upper (lower) arm of the interferometer is $z$ independent and is given by
\begin{align}\label{MZ-Ampl}
\begin{cases}
 {f}_{\text{up},\omega}^\text{MZI} = \sigma_\text{dx} {f}_{-,\omega}^\text{(1)}  +i\kappa_\text{dx} {f}_{-,\omega}^\text{(2)},  \\
 {f}_{\text{lo},\omega}^\text{MZI} = \left[i \kappa_\text{dx}  {f}_{-,\omega}^\text{(1)}  + \sigma_\text{dx} {f}_{-,\omega}^\text{(2)}\right]e^{i\Delta \phi}   \ ,
\end{cases}
\end{align}
where $f_{-,\omega}^{(i)}$ is the field enhancement of the $i^{th}$ resonator, defined in Eq. \eqref{FEmax}, and the slowly varying envelope functions at the end of the MZI in Resonator 1 and Resonator 2 are given by
\begin{align}
    f^\text{In}_{+,\omega} = &\left[ \sigma_{sx} \sigma_{dx}- \kappa_{sx}\kappa_{dx}e^{i\Delta \phi}\right]f^\text{In}_{-,\omega} \nonumber \\
    & + i\left[ \sigma_{sx}\kappa_{dx} + \kappa_{sx}\sigma_{dx}e^{i\Delta \phi} \right]f^\text{Add}_{-,\omega} 
    \end{align}
    and
    \begin{align}
    f^\text{Add}_{+,\omega} = &\left[- \kappa_{sx}\kappa_{dx} + \sigma_{sx} \sigma_{dx}e^{i\Delta \phi} \right] f^\text{Add}_{-,\omega} \nonumber \\
    & + i\left[ \kappa_{sx}\sigma_{dx} + \sigma_{sx}\kappa_{dx}e^{i\Delta \phi} \right]f^\text{In}_{-,\omega} \ ,
\end{align}
respectively. We note that, unlike for the DC structure, $f^\text{(1,2)}_{+,\omega} \neq {f}_{\text{up(lo)},\omega}^\text{MZI}(L_\text{MZI})$ and $f^\text{(1,2)}_{-,\omega} \neq {f}_{\text{up(lo)},\omega}^\text{MZI}(0)$ because of the field discontinuity introduced at each point coupler.  

Although the linear uncoupling provided by the MZI is not sensitive to the splitting ratio of the PCs, as long as they are identical, this parameter plays an important role when we focus on maximizing the nonlinear interaction. Considering the symmetry of the structure, if we assume we work in a narrow enough frequency band that the coupling can be considered frequency independent, the best configuration would be to work with PCs with a splitting ratio of 50:50. This would lead to a splitting of both pump and generated fields in both arms of the MZI, maximizing the overlap of the fields and thus the nonlinear interaction. If, instead, the splitting ratio were 100:0, the pump would be confined in the upper arm, while the signal and idler would be in the lower arm of the interferometer. Similarly, if the splitting ratio were 0:100, the situation would be the reverse. Thus, in both cases the nonlinear interaction would vanish.
In a more complicated situation, with the pump, signal, and idler at very different frequencies, the frequency dependence of the PCs could not be ignored, and one would have to design the structure accounting for different coupling ratios for the different fields, with the goal of directing them to
the same arm of the interferometer.

One can identify two advantages of the MZI coupler over the DC.  The first is that the performance of the MZI coupler is less affected than that of the DC by the frequency dependence of the coupling coefficients, and thus of the splitting ratio, since perfect linear uncoupling holds for the MZI coupler with identical PCs. The fabrication of a device with sufficiently identical couplers was discussed \cite{sabattoli2022nonlinear}, where linear uncoupling over a bandwidth of the order of hundreds of nanometers was achieved in the telecom band.
The second advantage is a higher photon conversion efficiency for the MZI coupler than for the DC; in the MZI the field distribution in each arm of the interferometer is the same as that of an isolated channel, and thus the slowly varying envelope function component is not oscillating.

\section{Comparing SFWM generation rates}
\label{SectionGeneration}

\begin{figure}[b]
\includegraphics[width=0.48\textwidth]{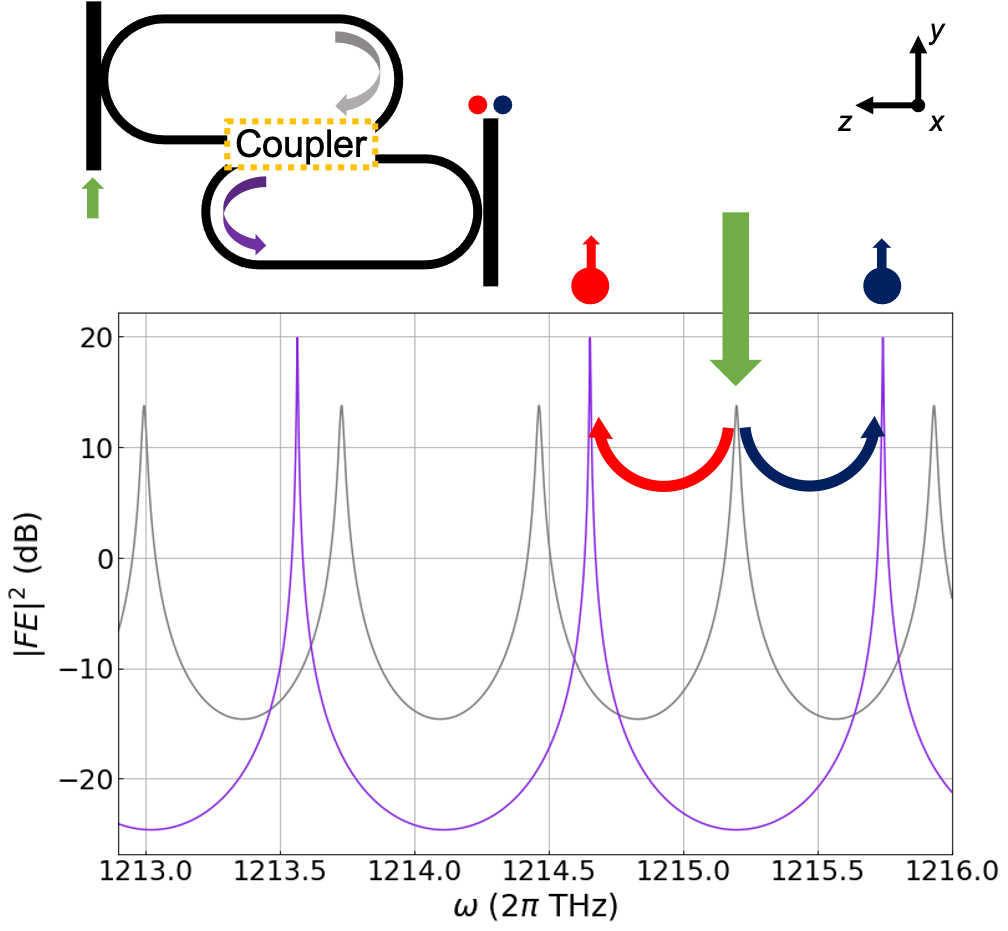}
\caption{\label{fig:SFWM} Resonances involved in the SFWM process. The green arrow represents the coherent pump that generates the pair of signal and idlers photons (red and blue dots, respectively). The resonances of the upper resonator are plotted in gray and those of the lower one are in violet. The interaction region, highlighted with the yellow dashed box in the structure's sketch, coincides with the coupling region.}
\end{figure}

In this section we study the generation of photon pairs by SFWM. The schematic idea of the process is shown in Fig. \ref{fig:SFWM}: a strong coherent pump is injected in the upper resonator (on resonance with one of its modes at $\omega_P$), and the signal-idler pair is generated on two resonances of the lower resonator at frequencies $\omega_S$ and $\omega_I$ and then is collected out of it. 
Since we are considering SFWM involving modes belonging to two different resonators, the only region of the structure that contributes to the nonlinear interaction is the coupler, highlighted with the dotted yellow box.
In order to make the nonlinear process possible, the energy position of the resonances of pump (gray solid line) and signal and idler (violet dashed line) must satisfy energy conservation.
The necessary fine control on the relative position of the two combs of resonances is achieved thanks to the linear uncoupling strategy.

The couplers that are the focus of this paper -- the DC and the MZI coupler -- fall in the category of couplers composed of two channels. Within this general framework we take $z$ to identify the propagation direction, with $x$ and $y$ being the transverse coordinates. Crucial to the calculations is the description of the overlap of the fields in the coupling region, identified by the overlap integral.  This quantity, which depends on the structure of the coupling region, plays a central role in determining the efficiency of the nonlinear process.  In the following we assume to good approximation a factorization of the overlap integral into a sort of effective area $\text{S}_\perp(\omega)$ and a spatial integral $\mathcal{J}(\omega)$.
This is discussed explicitly in Appendix \ref{app:GenRate}, in which the expression of the biphoton wavefunction describing the generated photon pairs is derived.

If we make the undepleted pump approximation and work in a low gain regime where there is a low probability of photon pair generation ($|\beta|^2 \ll 1$), we find the expression for the number of generated pair per pump pulse,

\begin{align}\label{Beta}
&|\beta|^2 = \frac{ \hbar^2 |\alpha|^4}{8 \pi^2} \frac{\gamma_\text{NL}^2}{\omega_P^2} \int d\omega_1 d\omega_2 \, \omega_1 \omega_2  \nonumber \\
& \times \left|  \int d\omega_3 \phi_P(\omega_3) \phi_P(\omega_4)    \sqrt{\omega_3 \omega_4} \, \mathcal{J}(\omega_1, \omega_2, \omega_3, \omega_4)   \right|^2  
\end{align}
(see Appendix \ref{app:GenRate} for a detailed calculation), where $|\alpha|^2$ is the average  number  of  pump  photons per pulse, $\gamma_\text{NL}$ is the nonlinear power factor, $\phi_{P}(\omega)$ is the pump profile, and

\begin{align}\label{Joverlap}
\mathcal{J}(\omega_1,\omega_2,\omega_3,\omega_4) = \sum_{ch} \mathcal{J}_{ch}(\omega_1,\omega_2,\omega_3,\omega_4) 
\end{align}
is the spatial integral of the $z$-dependent functions of the four fields involved in the process, where 

\begin{align}\label{Jch}
&\mathcal{J}_{ch}(\omega_1,\omega_2,\omega_3,\omega_{4}) =  \nonumber \\ 
&\int_0^{L_{ch}}  dz \, {f}_{ch,\omega_{1}}^*(z) {f}_{ch,\omega_{2}}^*(z) {f}_{ch,\omega_{3}}(z) {f}_{ch,\omega_{4}}(z) e^{i \Delta k z} 
\end{align}
is the integral in each channel, with $\Delta k = k(\omega_1)+k(\omega_2)-k(\omega_3)-k(\omega_4)$.
The shape of the field enhancement ${f}_{ch,\omega}(z)$ depends on the coupler under consideration. 
In general, eq. \eqref{Beta} needs to be evaluated numerically, but in order to calculate an explicit expression we work in the continuous wave (cw) regime by taking a narrow pump pulse $\phi_P(\omega)$ (see \cite{onodera2016parametric} for details), obtaining the photon-pair-generation rate

\begin{align}\label{R}
R_\text{pair} = \frac{|\beta|^2}{\Delta T} =& \frac{ 1}{4 \pi} \left(\frac{\gamma_\text{NL} P_P}{\omega_P} \right)^2 \int d\omega_1 \omega_1 (2\omega_P - \omega_1) \nonumber \\
& \times \left|  \mathcal{J}(\omega_1, 2\omega_P - \omega_1, \omega_P, \omega_P)  \right|^2 \ ,
\end{align}
where $P_P = \hbar \omega_P |\alpha|^2 / \Delta T$ is the injected pump power and $\Delta T$ is the pump temporal duration, taken to go to infinity along with the average number of pump photons in the pulse so that $P_P$ is held constant. We can estimate the pair-production rate for each structure once the expression of the overlap integral is evaluated.
Again, in general the integral in \eqref{R} needs to be evaluated numerically, but with the approximation of the Lorentzian shape of the intensity enhancement, described by eq. \eqref{ZingFE}, we can write an analytic expression for the pair generation rate.
In fact, with this approximation the integral \eqref{Jch} for a single channel can be written as

\begin{align}\label{JchLor}
|\mathcal{J}(\omega_1,&2\omega_P-\omega_1,\omega_P,\omega_P)|^2  \nonumber \\ 
& = |\mathcal{F}_\text{max}|^2 \mathcal{L}(\omega_1,2\omega_P-\omega_1)  |\mathcal{J}_\text{spatial}|^2 \ ,
\end{align}
where 

\begin{align}
    |\mathcal{F}_\text{max}|^2 = |\text{FE}_{(\text{max})2,S}|^2 |\text{FE}_{(\text{max})2,I}|^2 |\text{FE}_{(\text{max})1,P}|^4 
\end{align}
is the product of the maximum values of the intensity enhancement of each resonance involved in the process, given by eq. \eqref{FEmax}, and
\begin{align}\label{Lorentzians}
\mathcal{L}(\omega_1&,2\omega_P-\omega_1)  \nonumber \\
& = \left[ \frac{\frac{\Gamma_{2,S}^2}{4}}{\frac{\Gamma_{2,S}^2}{4}+(\omega_1-\omega_{S})^2}
\frac{\frac{\Gamma_{2,I}^2}{4}}{\frac{\Gamma_{2,I}^2}{4}+(\omega_1-\omega_{S})^2} \right]  \nonumber \\
& + \left[ \frac{\frac{\Gamma_{2,S}^2}{4}}{\frac{\Gamma_{2,S}^2}{4}+(\omega_1-\omega_{I})^2}
\frac{\frac{\Gamma_{2,I}^2}{4}}{\frac{\Gamma_{2,I}^2}{4}+(\omega_1-\omega_{I})^2} \right]
\end{align}
is the product of the Lorentzian shape of the signal and idler resonances centered at $\omega_S$ and $\omega_I$, respectively. Here $\Gamma_{2,m=S,I}$ is given by eq. \eqref{Gamma}, while the absolute value squared of the pump shape is assumed to be proportional to a Dirac $\delta$ function, given the cw regime; $\mathcal{J}_\text{spatial}$ is the spatial part of the integral, which depends only on the structure of the system.
We can perform the integral over the frequency-dependent factor, common for all the possible geometries, which results in

\begin{align}\label{LorIntegral}
\int & d\omega_1 \omega_1 (2\omega_P - \omega_1) \mathcal{L}(\omega_1,2\omega_P-\omega_1)  \nonumber \\ 
& \approx
\pi \frac{\Gamma_{2,S}\Gamma_{2,I}}{\left(\Gamma_{2,S}+\Gamma_{2,I}\right)}\left(\omega_S\omega_I\right)
\end{align}
(see Appendix \ref{app:LorentiansIntegral}).
With this in hand, we can write \eqref{R} as

\begin{align}\label{Ran}
R_\text{pair} =& \frac{1}{4 \pi} \left(\frac{\gamma_\text{NL} P_P}{\omega_P} \right)^2 |\mathcal{F}_\text{max}|^2 \nonumber \\ 
& \times \pi \frac{\Gamma_{2,S}\Gamma_{2,I}}{\left(\Gamma_{2,S}+\Gamma_{2,I}\right)}\left(\omega_S\omega_I\right) |\mathcal{J}_\text{spatial}|^2 \ .
\end{align}
The expression for $\mathcal{J}_\text{spatial}$ needs to be evaluated for each structure. In the following this we do and then compare our results for the DC and MZI coupling structures with that of a standard ring resonator, assuming in all calculations that we satisfy the phase-matching condition, $\Delta k = 0$, to good approximation.

\subsection{Ring resonator}
In the case of a simple ring resonator of length $L = 2 \pi R$, the spatial integral is 

\begin{align} \label{ZOverlapRing}
&\mathcal{J}_\text{spatial}^{\text{ring}}  = \int_0^L  e^{i \Delta k z} dz = L e^{i\frac{\Delta k L}{2} } \text{sinc}\left(  \frac{\Delta k L}{2}  \right)
\approx L  \ ,
\end{align} 
and hence the generation rate \eqref{Ran} is

\begin{align}\label{RpairRing}
R_\text{pair}^{\text{ring}} = & 
|\text{FE}_{(\text{max}),S}|^2 |\text{FE}_{(\text{max}),I}|^2 |\text{FE}_{(\text{max}),P}|^4
\nonumber \\
& \times \left(\frac{\gamma_{\text{NL}} P_P}{\omega_P}\right)^2 
\frac{\Gamma_{S}\Gamma_{I}}{\left(\Gamma_{S}+\Gamma_{I}\right)} \left(\omega_S\omega_I\right) \frac{L^2}{4} \ ,
\end{align}
with $\Gamma_{m=S,I}$ being the linewidth of the signal and idler resonances.
This can be expressed as a function of the loaded and coupling  quality factors ($Q_m$ and $Q_{C,m}$, respectively) using $\Gamma_m = \omega_m / Q_{m}$ and 
\begin{align}\label{IEprl}
	|\text{FE}_{(\text{max}),m}|^2 = \frac{1-\sigma_m^2}{(1-\sigma_m a_m)^2} = \frac{4 v_g}{L \omega_m} \frac{Q_{m}^2}{Q_{C,m}} \ ,
\end{align}
from \cite{menotti2019nonlinear}, giving
\begin{align}\label{RpairRingQ}
R_\text{pair}^{\text{ring}} = & \left(\frac{\gamma_{\text{NL}} P_P}{\omega_P}\right)^2  \left(  \frac{4 v_g}{L} \right)^4   \frac{  Q_{P}^4Q_{S}^2Q_{I}^2}{Q_{C,P}^2Q_{C,S}Q_{C,I}} \nonumber \\
& \times \frac{\omega_S\omega_I}{\omega_P^2}
\frac{L^2/4}{\omega_S Q_I +\omega_I Q_S}  \nonumber \\
= & \frac{4^3 \gamma_{\text{NL}}^2 P_P^2 v_g^4\omega_S \omega_I}{L^2\omega_P^4\left(\omega_S Q_I +\omega_I Q_S\right)}    \frac{  Q_{P}^4Q_{S}^2Q_{I}^2}{Q_{C,P}^2Q_{C,S}Q_{C,I}} \ .
\end{align}
We assume we are working over a small enough frequency range that we can take  $\omega_P \approx \omega_S \approx \omega_I = \omega$ and $Q_{P} \approx Q_{S} \approx Q_{I}=Q$, which allows us to reduce \eqref{RpairRing} to a simplified form. For no losses ($Q_{C,m}=Q_{m}$), we find
\begin{align}\label{RpairNoloss}
R_\text{pair}^{\text{ring}} & =  \left(\gamma_{\text{NL}} P_P\right)^2 \frac{32 v_g^4}{\omega^3}\frac{Q^3}{L^2} \ ,
\end{align}
retrieving the SFWM generation rate expression found earlier \cite{helt2012does}, while for critical coupling ($Q_{C,m}=2 Q_{m}$) we find
\begin{align}\label{RpairCC}
R_\text{pair}^{\text{ring}} & =  \left(\gamma_{\text{NL}} P_P\right)^2 \frac{2 v_g^4}{\omega^3}\frac{Q^3}{L^2} \ .
\end{align}

\subsection{Directional coupler}

If we consider the DC structure and assume perfect uncoupling in the linear regime, the spatial integral is in the form of (a detailed calculation is given in Appendix \ref{app:Jspatial})

\begin{align} \label{ZOverlapDC}
\mathcal{J}_\text{spatial}  \approx - \frac{L_\text{DC}}{4}\left[  1 - \text{sinc}\!\left(4 \kappa_\text{DC}  L_\text{DC}\right)\right]  \ ,
\end{align} 
which lead to a generation rate of 

\begin{align}\label{RpairZing}
&R_\text{pair}^\text{DC}   \nonumber \\
& = \left(\frac{\gamma_{\text{NL}} P_P}{\omega_P}\right)^2 
|\text{FE}_{(\text{max})2,S}|^2 |\text{FE}_{(\text{max})2,I}|^2 |\text{FE}_{(\text{max})1,P}|^4 \nonumber \\
& \times \frac{\Gamma_{2,S}\Gamma_{2,I}}{\Gamma_{2,S}+\Gamma_{2,I}}\left(\omega_S \omega_I \right) 
\frac{L_\text{DC}^2}{64}\left[  1 - \text{sinc}\!\left(4 \kappa_\text{DC}  L_\text{DC}\right)\right]^2  \ .
\end{align}
Again, we can express this formula in terms of quality factors, and find
\begin{align}\label{RpairZingQ}
R_\text{pair}^\text{DC} = \, &\frac{4 \gamma_{\text{NL}}^2 P_P^2 v_g^4\omega_S \omega_I}{\omega_P^4\left(\omega_S Q_I +\omega_I Q_S\right)}    \frac{Q_{P}^4Q_{S}^2Q_{I}^2}{Q_{C,P}^2Q_{C,S}Q_{C,I}} \nonumber \\
&\times \left(  \frac{L_\text{DC} }{L_1 L_2} \right)^2
\left[ 1 - \text{sinc}\!\left(4 \kappa_\text{DC} L_\text{DC} \right)\right]^2 \ .
\end{align}

\subsection{Mach-Zehnder interferometer}

For the MZI coupler structure with perfect uncoupling in the linear regime, taking $\sigma_\text{dx} = \sigma_\text{sx} =1/\sqrt{2}$, i.e., 50:50 beam splitters, the spatial integral results in 

\begin{align} \label{ZOverlapMZI}
&\mathcal{J}_\text{spatial} \approx - \frac{L_\text{MZI}}{2}   \ ,
\end{align} 
(a detailed calculation is given in Appendix \ref{app:Jspatial}), which leads to a generation rate of 

\begin{align}\label{RpairMZing}
R_\text{pair}^\text{MZI}= & \left(\frac{\gamma_{\text{NL}} P_P}{\omega_P}\right)^2 
|\text{FE}_{(\text{max})2,S}|^2 |\text{FE}_{(\text{max})2,I}|^2 \nonumber \\
& \times |\text{FE}_{(\text{max})1,P}|^4 
\frac{\Gamma_{2,S}\Gamma_{2,I}}{\Gamma_{2,S}+\Gamma_{2,I}}\left(\omega_S \omega_I \right) 
\frac{L_\text{MZI}^2}{16} \ ,
\end{align}
and in terms of quality factors 
\begin{align}\label{RpairMZingQ}
R_\text{pair}^\text{MZI} = \, &  \frac{16 v_g^4 \gamma_{\text{NL}}^2 P_P^2 \omega_S \omega_I}{\omega_P^4 \left( \omega_S Q_I + \omega_I Q_S \right) } \nonumber \\
& \times \frac{Q_{P}^4Q_{S}^2Q_{I}^2}{Q_{C,P}^2Q_{C,S}Q_{C,I}}  \left(  \frac{L_\text{MZI} }{L_1 L_2} \right)^2 \ .
\end{align}

\subsection{Comparison}

We can have better insight into the generation efficiency of our two structures if we directly compare it with the ring resonator.
For the DC structure, the sinc function in \eqref{RpairZing} makes a negligible contribution for our typical parameters of interest, and we can write 
\begin{align}\label{RcomparisonZing1}
R_\text{pair}^\text{DC}\approx & \left(\frac{L L_\text{DC}}{4 L_1 L_2}\right)^2  R_\text{pair}^\text{ring} \ ,
\end{align}
and assuming $L_1 = L_2 = 2L$, which means having two racetracks with the same bending radius of the ring, and taking the optimal length of the DC, i.e., $L_\text{DC}=\pi R$ as shown in \cite{menotti2019nonlinear}, the expression becomes
\begin{align}\label{RcomparisonZing2}
R_\text{pair}^\text{DC} \approx  \frac{1}{1024} R_\text{pair}^\text{ring}\ .
\end{align}
We can do the same calculations for the MZI structure,
\begin{align}\label{RcomparisonMZing1}
R_\text{pair,ext}^\text{MZI}\approx & \left(\frac{L L_\text{MZ}}{2 L_1 L_2}\right)^2  R_\text{pair}^\text{ring}
\ ,
\end{align}
and assuming $L_1=L_2 = 2L$ and $L_\text{MZ}=\pi R$,
\begin{align}\label{RcomparisonMZing2}
R_\text{pair}^\text{MZI} \approx \frac{1}{256} R_\text{pair}^\text{ring} \ .
\end{align}
An overview of our estimates is reported in Table \ref{tbl:RateComparison}, along with the main quantities of interest of each structure. Although these two rates are substantially reduced from that of the standard ring, the significant benefit obtained is that we can implement independent control on each comb of resonances, and achieve several decibels of pump filtering from the generated pair.

\section{Conclusions}

In this work we considered structures composed of two linearly uncoupled racetrack resonators, in which energy transfer between them can occur only through a nonlinear interaction. We considered two strategies to uncouple the racetracks: via the use of a directional coupler and via the use of a Mach-Zehnder interferometer. 
The DC approach is simple and can be realized in compact structures \cite{sabattoli2021}, but it has two main drawbacks: first, its properties can be considered frequency independent only over a limited bandwidth of typically a few tens of nanometers at telecom wavelengths; second, the fields inside the coupler oscillate, reducing the nonlinear interaction efficiency.
On the other hand, the MZI approach guarantees linear uncoupling isolation 
over a much larger bandwidth (hundreds of nanometers \cite{sabattoli2022nonlinear}, thanks to the much lower frequency-sensitivity of the interference mechanism, and offers a higher conversion efficiency, given that the slowly-varying envelope functions of the fields do not oscillate. The situation where the fields are at very distant frequencies, and thus the coupling ratios cannot be considered frequency-independent, will be addressed in future work.

The particular nonlinear interaction we considered was spontaneous four-wave mixing, which can be exploited for the generation of photon pairs. We developed the theory to describe such a quantum nonlinear process in the kind of structures we studied; it is more complicated than the theory for the single ring resonator more commonly studied. Although the generation rates are lower than that of a single ring, we showed how the DC and MZI coupler structures we studied can be used to control the generation of quantum correlated photon pairs from two perspectives.
First, we can fine tune the resonances of the resonators, and hence the frequency of the photons generated, independently. Second, the structures
also provide the filtering of the pump from the signal and idler photons, necessary when working with SFWM process. Explicit expressions for the efficiency of the pair generation rate for the different structures, together with an expression for the biphoton wave function, were given. 

\section*{Acknowledgments}
M.L. and L.Z. acknowledge support from Ministero dell’Istruzione, dell’Università e della Ricerca (Dipartimenti di Eccellenza 20182022 F11I18000680001). J.E.S. acknowledges support from the Natural Sciences and Engineering Research Council of Canada.

\newpage 

\widetext

\begin{table}[b]
\begin{center}
\begin{tabular}{ | c | c | c | c |}
    \hline
    Structure & Spatial integral & Overall finesse factor & Pair-generation rate \\ 
    \hline
    \parbox[c]{4cm}{\includegraphics[width=0.1\textwidth]{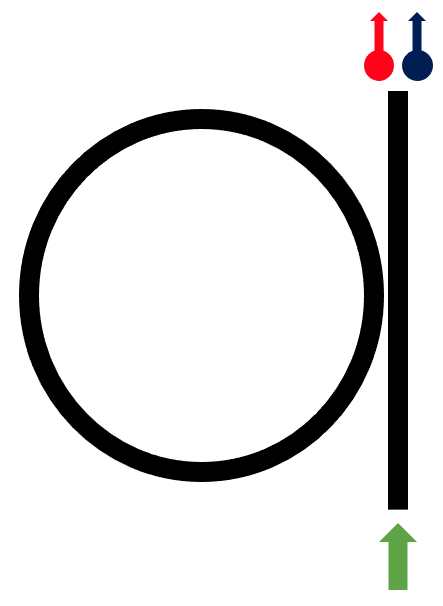} }
    &
    \parbox[c]{4cm}{\begin{equation*}
                    \mathcal{J}_\text{spatial}^{\text{ring}} \approx L 
                    \end{equation*}}
    &
    \parbox[c]{3cm}{\begin{equation*}
                    \mathcal{F} \propto \left(\frac{1}{L}\right)^4  
                    \end{equation*}}
    &
    \parbox[c]{5.5cm}{\begin{align*}
                    R_\text{pair}^\text{ring} = &
                    \frac{4^3 \Gamma^2 P_P^2 v_g^4\omega_S \omega_I}{L^2\omega_P^4\left(\omega_S Q_I +\omega_I Q_S\right)}   \nonumber \\ 
                    & \times\frac{ Q_{P}^4Q_{S}^2Q_{I}^2}{Q_{C,P}^2Q_{C,S}Q_{C,I}}
                    \end{align*}} \eqref{RpairRingQ}
    \\
    \hline
    \parbox[c]{4cm}{\includegraphics[width=0.2\textwidth]{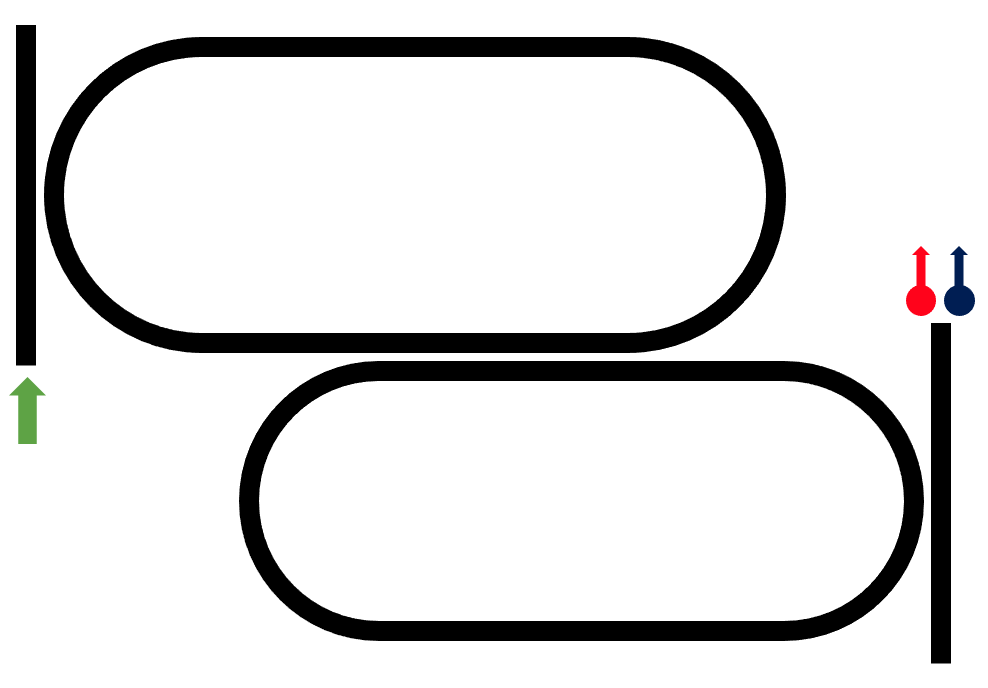} } 
    &
    \parbox[c]{4cm}{\begin{equation*}
                    \mathcal{J}_\text{spatial}^\text{DC} \approx \frac{L_\text{DC}}{4} 
                    \end{equation*}}
    &
    \parbox[c]{3cm}{\begin{equation*}
                    \mathcal{F} \propto \left(\frac{1}{L_1L_2}\right)^2  
                    \end{equation*}}
    &
    \parbox[c]{5.5cm}{\begin{equation*}
                    R_\text{pair}^\text{DC} \approx  \frac{1}{1024} R_\text{pair}^\text{ring} 
                    \end{equation*}} \eqref{RcomparisonZing2}
    \\
    \hline
    \parbox[c]{4cm}{\includegraphics[width=0.2\textwidth]{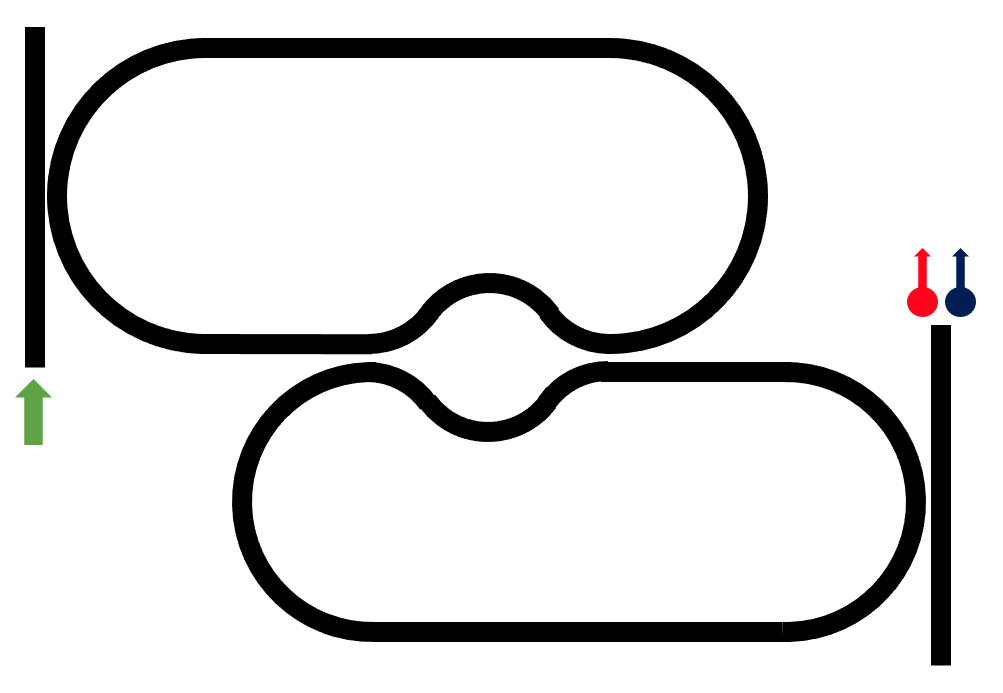} } 
    &
        \parbox[c]{4cm}{\begin{equation*}
                    \mathcal{J}_\text{spatial}^\text{MZI} \approx \frac{L_\text{MZI}}{2} 
                    \end{equation*}}
    &
    \parbox[c]{3cm}{\begin{equation*}
                    \mathcal{F} \propto \left(\frac{1}{L_1L_2}\right)^2 
                    \end{equation*}}
    &
    \parbox[c]{5.5cm}{\begin{equation*}
                    R_\text{pair}^\text{MZI} \approx  \frac{1}{256} R_\text{pair}^\text{ring} 
                    \end{equation*}} \eqref{RcomparisonMZing2}
    \\
    \hline
\end{tabular}
\caption{Rate comparison between the ring resonator, the DC resonator and the MZI resonator. The other quantities of interest are the spatial integral and the resonators' finesses.}
\label{tbl:RateComparison}
\end{center}
\end{table}

\twocolumngrid


\appendix

\section{Nonlinear Hamiltonian and generation rate}\label{app:GenRate}

In order to study the generation of photons pairs by parametric fluorescence, we follow the backward Heisenberg approach introduced earlier
\cite{yang2008spontaneous} to describe spontaneous parametric down-conversion; the same approach can be generalized to describe SFWM. In this method one starts from the Hamiltonian describing the electromagnetic field in the structure and uses it to define the evolution of the state. 
The third-order nonlinear interaction is controlled by a nonlinear Hamiltonian in the form of

\begin{equation} \label{Hnl}
\mathcal{H}_\text{NL} = -\frac{1}{4 \varepsilon_0} \int d\textbf{r} \Gamma_3^{ijlm}(\textbf{r}) D^i(\textbf{r}) D^j(\textbf{r}) D^l(\textbf{r}) D^m(\textbf{r}) \  ,
\end{equation}
where $\Gamma_3^{ijlm}(\textbf{r})$ is the third-order susceptibility tensor; 
$\textbf{D}(\textbf{r})$ are operators associated with the pump, signal, and idler fields, which depend on the structure under consideration, and $i,j,l,$ and $m$ are the Cartesian coordinates.
We follow a description (and quantization) of the electromagnetic field in terms of asymptotic fields \cite{liscidini2012asymptotic}. This approach is convenient, for it allows one to treat almost any geometry.
The expression for the mode fields $\textbf{D}(\textbf{r})$ in the asymptotic-field framework, recalling that we are neglecting GVD, is 
\begin{equation}\label{Asy}
\textbf{D}(\textbf{r}) = \int_0^\infty  d\omega \, \sqrt{\frac{\hbar \omega}{2 v_{g}}} a(\omega) \textbf{D}_\omega^\text{asy-in(out)}(\textbf{r}) + \text{H.c.} \  ,
\end{equation}
where we consider only one transverse mode, $a(\omega)$ is the destruction operator that destroys a photon at frequency $\omega$, and $\textbf{D}_\omega^\text{asy-in(out)}(\textbf{r})$ is the displacement vector associated with the asymptotic-in(-out) field, already described in eq. \eqref{Displacement}.
If we use \eqref{Asy} in \eqref{Hnl}, considering only operators that destroy two pump photons, $a_P(\omega_3)$ and $a_P(\omega_4)$, and creating the signal and idler photons, $a_S^\dagger(\omega_1)$ and $a_I^\dagger(\omega_2)$, respectively (see \cite{sipe2004effective} for details), we can rewrite the nonlinear Hamiltonian in a form that describes only the four-wave mixing process 

\begin{align}\label{Hnl2}
\mathcal{H}&_\text{NL}^\text{FWM} = - \int  d\omega_1 d\omega_2 d\omega_3 d\omega_{4} \text{S}_\perp(\omega_1,\omega_2,\omega_3,\omega_{4})  \nonumber \\
&\times \mathcal{J}(\omega_1,\omega_2,\omega_3,\omega_{4})
a^\dagger_{S}(\omega_1) a^\dagger_{I}(\omega_2) a_{P}(\omega_3) a_{P}(\omega_4)  \ ,
\end{align}
where 
\begin{align}\label{Sperp}
\text{S}_\perp&(\omega_1,\omega_2,\omega_3,\omega_{4})\nonumber \\
&  = \frac{3 \hbar^2}{(4\pi)^2\varepsilon_0} \sqrt{ \frac{\omega_1\omega_2\omega_3\omega_4}{v_g^4} } \int  dx dy \, \Gamma_3^{ijlm}(x,y)\nonumber \\
& \times \left[ \text{d}^i_{ch}(x,y)   \text{d}^j_{ch}(x,y)  \right]^* \text{d}^l_{ch}(x,y) \text{d}^m_{ch}(x,y)
\end{align}
is a nonlinear coupling term, assumed to be equal for both channels, which sums up the effective area including the integral over the transverse coordinates ($x,y$) and which can be expressed in terms of the nonlinear power factor \cite{onodera2016parametric}
\begin{align}\label{gammanl}
\gamma_{\text{NL}} = &\frac{3 \omega_P}{4 \varepsilon_0^3 v_{g}^2} \int dx dy \frac{\chi_3^{ijkl}(x,y)}{ n^8(x,y)} \times \nonumber \\
&\left[ \text{d}^i_{ch}(x,y)   \text{d}^j_{ch}(x,y)  \right]^* \text{d}^l_{ch}(x,y) \text{d}^m_{ch}(x,y) \ ,
\end{align}
where we assumed $\Gamma_3$ to be $z$ independent and expressed it in terms of the more familiar $\chi_3$, resulting in
\begin{align}\label{Sperpgamma}
&\text{S}_\perp(\omega_1,\omega_2,\omega_3,\omega_{4}) =  \frac{\hbar^2\gamma_\text{NL}}{4 \pi^2} \frac{\sqrt{\omega_1\omega_2\omega_3\omega_4}}{\omega_P}\ ,
\end{align}
where $ \mathcal{J}(\omega_1,\omega_2,\omega_3,\omega_4) $
is the integral of the $z$-dependent functions of the fields in Eq. \eqref{Joverlap}.
We have been able to separate the components $x$ and $y$ from $z$, thus factoring expression \ref{Hnl2} to involve the functions $\text{S}_\perp(\omega_1,\omega_2,\omega_3,\omega_{4})$ and $\mathcal{J}(\omega_1,\omega_2,\omega_3,\omega_4)$, only because we assume a polarization for the field pointing off the chip.
We make the undepleted pump approximation, since we assume the intensity of the generated photons is much smaller than the intensity of the pump, and treat the pump classically by taking $a_{P}(\omega) = \alpha \phi(\omega)$, where $|\alpha|^2$ is the average number of pump photons per pulse and $\phi_{P}(\omega)$ is the pump profile.
With these assumptions, eq. \eqref{Hnl2} leads to the expression for the biphoton wavefunction of the signal and idler pair 

\begin{align}\label{Biphoton}
&\phi(\omega_1,\omega_2) =  \frac{i \sqrt{2} \hbar \alpha^2}{4 \pi \beta} \frac{\sqrt{\omega_1 \omega_2}}{\omega_P}  \gamma_\text{NL}  \nonumber \\
& \times \int d\omega_3  \phi_P(\omega_3) \phi_P(\omega_4) \sqrt{\omega_3 \omega_4}
\, \mathcal{J}(\omega_1, \omega_2, \omega_3, \omega_4)   \ ,
\end{align}
where $\omega_4 = \omega_1 + \omega_2 - \omega_3$ is set by the energy conservation and $\omega_3$ is the pump photon frequency over which the integral is performed.
From the normalization of the biphoton wavefunction, $\int d\omega_1 d\omega_2| \phi(\omega_1,\omega_2)|^2 = 1$, we obtain the number of generated pair per pulse
of Eq. \eqref{Beta}. Alternatively, those rates can be identified directly by a calculation using Fermi's golden rule \cite{banic2022two}.

\section{Analytic calculation of the Lorentzian shape of the resonances}
\label{app:LorentiansIntegral}
Equation \eqref{Lorentzians} is calculated from the Lorentzian shape of signal and idler resonances,

\begin{align}
\ell(\omega_1) = \left[ \frac{\frac{\Gamma_{2,S}^2}{4}}{\frac{\Gamma_{2,S}^2}{4}+(\omega_1-\omega_{S})^2}+
\frac{\frac{\Gamma_{2,I}^2}{4}}{\frac{\Gamma_{2,I}^2}{4}+(\omega_1-\omega_{I})^2} \right] 
\end{align}
and
\begin{align}
&\ell(2 \omega_P -\omega_1) = 
\left[ \frac{\frac{\Gamma_{2,S}^2}{4}}{\frac{\Gamma_{2,S}^2}{4}+(2 \omega_P -\omega_1-\omega_{S})^2}\right. \nonumber \\
& \hspace{2.5cm} \left. + \frac{\frac{\Gamma_{2,I}^2}{4}}{\frac{\Gamma_{2,I}^2}{4}+(2 \omega_P -\omega_1-\omega_{I})^2} \right] \nonumber \\
& =  \left[ \frac{\frac{\Gamma_{2,S}^2}{4}}{\frac{\Gamma_{2,S}^2}{4}+(\omega_1-\omega_{I})^2}+
\frac{\frac{\Gamma_{2,I}^2}{4}}{\frac{\Gamma_{2,I}^2}{4}+(\omega_1-\omega_{S})^2} \right] \ ,
\end{align}
where we have considered signal and idler photons to be indistinguishable and thus that they both can be generated in the resonance around $\omega_{S}$ and in that around $\omega_{I}$. These give the expression
\begin{align}
 \mathcal{L}(\omega_1, & 2 \omega_P -\omega_1) = \ell (\omega_1) \ell (2 \omega_P -\omega_1)  \nonumber \\
& \approx\left[ \frac{\frac{\Gamma_{2,S}^2}{4}}{\frac{\Gamma_{2,S}^2}{4}+(\omega_1-\omega_{S})^2}
\frac{\frac{\Gamma_{2,I}^2}{4}}{\frac{\Gamma_{2,I}^2}{4}+(\omega_1-\omega_{S})^2} \right]  \nonumber \\
& + \left[ \frac{\frac{\Gamma_{2,S}^2}{4}}{\frac{\Gamma_{2,S}^2}{4}+(\omega_1-\omega_{I})^2}
\frac{\frac{\Gamma_{2,I}^2}{4}}{\frac{\Gamma_{2,I}^2}{4}+(\omega_1-\omega_{I})^2} \right]  \nonumber \\
& = \mathcal{C}_1(\omega_1) +  \mathcal{C}_2(\omega_1) \ ,
\end{align}
where we have neglected the product of Lorentzians centered at different frequencies, given that, generally, $\text{FSR}_i \gg \Gamma_{i,m}$. This leads to
\begin{align}
\int d\omega_1 & \omega_1 (2\omega_P - \omega_1) \mathcal{L}(\omega_1,2 \omega_P -\omega_1)  \nonumber \\
& = \int d\omega_1 \omega_1 (2\omega_P - \omega_1) \left[\mathcal{C}_1(\omega_1) +  \mathcal{C}_2(\omega_1)\right] \ ,
\end{align}
where
\begin{align}
\int d\omega_1 &\omega_1 (2\omega_P - \omega_1) \mathcal{C}_1(\omega_1) \nonumber \\
& = \pi \frac{ \frac{\Gamma_{2,S}}{2} \frac{\Gamma_{2,I}}{2} }{\frac{\Gamma_{2,S}^2}{4} - \frac{\Gamma_{2,I}^2}{4} } 
\left[ \frac{\Gamma_{2,S}}{2}\left(2\omega_P \omega_S+\frac{\Gamma_{2,I}^2}{4} -\omega_S^2\right) \right. \nonumber \\
& - \left. \frac{\Gamma_{2,I}}{2} \left(2\omega_P \omega_S+\frac{\Gamma_{2,S}^2}{4} -\omega_S^2\right)  \right]  \nonumber \\
& = \frac{\pi}{2} \frac{\Gamma_{2,S}\Gamma_{2,I}\left(2\omega_P \omega_S-\omega_S^2\right)}{\left(\Gamma_{2,S}+\Gamma_{2,I}\right)\left(\Gamma_{2,S}-\Gamma_{2,I}\right)}\left(\Gamma_{2,S}-\Gamma_{2,I}\right) \nonumber \\
& = \frac{\pi}{2} \frac{\Gamma_{2,S}\Gamma_{2,I}}{\left(\Gamma_{2,S}+\Gamma_{2,I}\right)}\left(2\omega_P \omega_S-\omega_S^2\right) \nonumber \\
& = \frac{\pi}{2} \frac{\Gamma_{2,S}\Gamma_{2,I}}{\left(\Gamma_{2,S}+\Gamma_{2,I}\right)}\left(\omega_S\omega_I\right) \ , 
\end{align}
where we took $\left(2\omega_P \omega_S+\frac{\Gamma_{2,m}^2}{4} -\omega_S^2\right)\approx \left(2\omega_P \omega_S - \omega_S^2\right)$ since $\Gamma_{2,m} \ll \omega_m^2$ and 

\begin{align}
\int d\omega_1 & \omega_1 (2\omega_P - \omega_1) \mathcal{C}_2(\omega_1) \nonumber \\
& = \frac{\pi}{2} \frac{\Gamma_{2,S}\Gamma_{2,I}}{\left(\Gamma_{2,S}+\Gamma_{2,I}\right)}\left(2\omega_P \omega_I-\omega_I^2\right) \nonumber \\
& = \frac{\pi}{2} \frac{\Gamma_{2,S}\Gamma_{2,I}}{\left(\Gamma_{2,S}+\Gamma_{2,I}\right)}\left(\omega_S\omega_I\right) \ , 
\end{align}
and hence to the final value
\begin{align}
\int d\omega_1 & \omega_1 (2\omega_P - \omega_1) \mathcal{L}(\omega_1,2 \omega_P -\omega_1) =  \nonumber \\
& = \pi \frac{\Gamma_{2,S}\Gamma_{2,I}}{\left(\Gamma_{2,S}+\Gamma_{2,I}\right)}\left(\omega_S\omega_I\right) \ , 
\end{align}
of Eq. \eqref{LorIntegral}.

\section{Analytic calculation of $\mathcal{J}$ spatial}
\label{app:Jspatial}
In the case of perfect uncoupling via the DC we have ${f}_{\text{up},\omega_{1(2)}}(0) = {f}_{\text{lo},\omega_{3(4)}}(0) = 0$, and hence, the condition \eqref{DC-Ampl} for each frequency simplifies to

\begin{align}
\begin{cases}
 {f}_{\text{up},\omega_{1(2)}}^{\text{DC}}(z)  =  - i{f}_{\text{lo},\omega_{1(2)}}(0) \sin(|\kappa_\text{DC}|z) \\
 {f}_{\text{up},\omega_{3(4)}}^{\text{DC}}(z)  =  {f}_{\text{up},\omega_{3(4)}}(0) \cos(|\kappa_\text{DC}|z) \\
 {f}_{\text{up},\omega_{1(2)}}^{\text{DC}}(z)  =  {f}_{\text{up},\omega_{1(2)}}(0) \cos(|\kappa_\text{DC}|z) \\
 {f}_{\text{up},\omega_{3(4)}}^{\text{DC}}(z)  =  - i{f}_{\text{lo},\omega_{3(4)}}(0) \sin(|\kappa_\text{DC}|z) \\ 
\end{cases}\ ,
\end{align}
giving, from \eqref{Jch},
\begin{align}
\mathcal{J}_{\text{up}}&(\omega_1,\omega_2,\omega_3,\omega_4) = \mathcal{J}_{\text{lo}}(\omega_1,\omega_2,\omega_3,\omega_4) \nonumber \\ 
& = {f}_{\text{lo},\omega_{1}}(0){f}_{\text{lo},\omega_{2}}(0) {f}_{\text{up},\omega_{3}}(0) {f}_{\text{up},\omega_{4}}(0) \nonumber \\ 
& \times \int_0^{L_{\text{DC}}}  i^2  \sin^2(|\kappa_\text{DC}|z)  \cos^2(|\kappa_\text{DC}|z) e^{i \Delta k z} dz
\end{align}
and hence 
\begin{align}
\mathcal{J}&(\omega_1,\omega_2,\omega_3,\omega_4) \nonumber \\ 
& = {f}_{\text{lo},\omega_{1}}(0){f}_{\text{lo},\omega_{2}}(0) {f}_{\text{up},\omega_{3}}(0) {f}_{\text{up},\omega_{4}}(0) \mathcal{J}_\text{spatial}^{\text{DC}} \ ,
\end{align}
with
\begin{align} 
\mathcal{J}_\text{spatial}^{\text{DC}} &= 2 \int_0^{L_\text{DC}}  - \cos^2(\kappa_\text{DC}z) \sin^2(\kappa_\text{DC}z)  e^{i \Delta k z} dz \nonumber \\
& \approx - \frac{L_\text{DC}}{4}\left[  1 - \text{sinc}\!\left(4 \kappa_\text{DC}  L_\text{DC}\right)\right]  \ .
\end{align} 

Similarly, in the case of perfect uncoupling via a MZI coupler, we have ${f}_{-,\omega_{1(2)}}^\text{(1)} = {f}_{-,\omega_{3(4)}}^\text{(2)} = 0$, and if we assume balanced beam splitters, $\sigma_\text{dx} = \sigma_\text{sx} = \sigma $ (and hence $\kappa_\text{dx} = \kappa_\text{sx} = \kappa$), condition \eqref{MZ-Ampl} for the different frequencies simplifies to

\begin{align}
\begin{cases}
 {f}_{\text{up},\omega_{1(2)}}^{\text{MZI}}(z)  =  i \kappa  {f}_{-,\omega_{1(2)}}^\text{(2)}  \\
 {f}_{\text{up},\omega_{3(4)}}^{\text{MZI}}(z)  = \sigma  {f}_{-,\omega_{3(4)}}^\text{(1)} \\
 {f}_{\text{lo},\omega_{1(2)}}^{\text{MZI}}(z)  = \sigma  {f}_{-,\omega_{1(2)}}^\text{(2)}  \\
 {f}_{\text{lo},\omega_{3(4)}}^{\text{MZI}}(z)  = i \kappa  {f}_{-,\omega_{3(4)}}^\text{(1)} \\
\end{cases}\ ,
\end{align}

and from \eqref{Jch},
\begin{align}
\mathcal{J}_{\text{up}}&(\omega_1,\omega_2,\omega_3,\omega_4) = \mathcal{J}_{\text{lo}}(\omega_1,\omega_2,\omega_3,\omega_4)  \nonumber \\ 
& = {f}_{-,\omega_{1}}^\text{(2)}{f}_{-,\omega_{2}}^\text{(2)} {f}_{-,\omega_{3}}^\text{(1)} {f}_{-,\omega_{4}}^\text{(1)} \int_0^{L_{\text{MZI}}}  i^2 \sigma^2 \kappa^2   e^{i \Delta k z} dz,
\end{align}
and hence,
\begin{align}
&\mathcal{J}(\omega_1,\omega_2,\omega_3,\omega_4) = {f}_{-,\omega_{1}}^\text{(2)}{f}_{-,\omega_{2}}^\text{(2)} {f}_{-,\omega_{3}}^\text{(1)} {f}_{-,\omega_{4}}^\text{(1)}  \mathcal{J}_\text{spatial}^\text{MZI}  \ ,
\end{align}
with
\begin{align} 
&\mathcal{J}_\text{spatial}^\text{MZI}   = 2 \int_0^{L_\text{MZI}} (- \sigma^2 \kappa^2 e^{i \Delta k z}) dz  
\approx - \frac{L_\text{MZI}}{2}   
\end{align} 
in the case of 50:50 point couplers $(\sigma=\kappa=1/\sqrt{2})$.

\section{Generation rates in terms of finesse}

The expression for the generation rate can be written in terms of the finesse $\mathcal{F}$ of the resonators, thanks to the relation found in Sec. II. For example, at the critical coupling condition, we have

\begin{align}\label{RpairRingFinesse}
R_\text{pair}^\text{ring}= & \left(\frac{\gamma_{\text{NL}} P_P}{\omega_P}\right)^2 
\left(\frac{\mathcal{F}_S}{\pi}\right) \left(\frac{\mathcal{F}_I}{\pi}\right) 
\left(\frac{\mathcal{F}_P}{\pi}\right)^2 \nonumber \\ 
& \times 
\frac{\Gamma_{2,S}\Gamma_{2,I}}{\Gamma_{2,S}+\Gamma_{2,I}}\left(\omega_S \omega_I \right) 
\frac{L^2}{4} \ ,
\end{align}

\begin{align}\label{RpairZingFinesse}
R_\text{pair}^\text{DC}= & \left(\frac{\gamma_{\text{NL}} P_P}{\omega_P}\right)^2 
\left(\frac{\mathcal{F}_S}{\pi}\right) \left(\frac{\mathcal{F}_I}{\pi}\right) 
\left(\frac{\mathcal{F}_P}{\pi}\right)^2 \nonumber \\ 
& \times 
\frac{\Gamma_{2,S}\Gamma_{2,I}}{\Gamma_{2,S}+\Gamma_{2,I}}\left(\omega_S \omega_I \right) 
\frac{L_\text{DC}^2}{64} \ ,
\end{align}
and
\begin{align}\label{RpairMachZingFinesse}
R_\text{pair}^\text{MZI}= & \left(\frac{\gamma_{\text{NL}} P_P}{\omega_P}\right)^2 
\left(\frac{\mathcal{F}_S}{\pi}\right) \left(\frac{\mathcal{F}_I}{\pi}\right) 
\left(\frac{\mathcal{F}_P}{\pi}\right)^2 \nonumber \\ 
& \times 
\frac{\Gamma_{2,S}\Gamma_{2,I}}{\Gamma_{2,S}+\Gamma_{2,I}}\left(\omega_S \omega_I \right) 
\frac{L_\text{MZI}^2}{16} \ .
\end{align}

\bibliography{Manuscript}

\begin{thebibliography}{32}%
\makeatletter
\providecommand \@ifxundefined [1]{%
 \@ifx{#1\undefined}
}%
\providecommand \@ifnum [1]{%
 \ifnum #1\expandafter \@firstoftwo
 \else \expandafter \@secondoftwo
 \fi
}%
\providecommand \@ifx [1]{%
 \ifx #1\expandafter \@firstoftwo
 \else \expandafter \@secondoftwo
 \fi
}%
\providecommand \natexlab [1]{#1}%
\providecommand \enquote  [1]{``#1''}%
\providecommand \bibnamefont  [1]{#1}%
\providecommand \bibfnamefont [1]{#1}%
\providecommand \citenamefont [1]{#1}%
\providecommand \href@noop [0]{\@secondoftwo}%
\providecommand \href [0]{\begingroup \@sanitize@url \@href}%
\providecommand \@href[1]{\@@startlink{#1}\@@href}%
\providecommand \@@href[1]{\endgroup#1\@@endlink}%
\providecommand \@sanitize@url [0]{\catcode `\\12\catcode `\$12\catcode
  `\&12\catcode `\#12\catcode `\^12\catcode `\_12\catcode `\%12\relax}%
\providecommand \@@startlink[1]{}%
\providecommand \@@endlink[0]{}%
\providecommand \url  [0]{\begingroup\@sanitize@url \@url }%
\providecommand \@url [1]{\endgroup\@href {#1}{\urlprefix }}%
\providecommand \urlprefix  [0]{URL }%
\providecommand \Eprint [0]{\href }%
\providecommand \doibase [0]{https://doi.org/}%
\providecommand \selectlanguage [0]{\@gobble}%
\providecommand \bibinfo  [0]{\@secondoftwo}%
\providecommand \bibfield  [0]{\@secondoftwo}%
\providecommand \translation [1]{[#1]}%
\providecommand \BibitemOpen [0]{}%
\providecommand \bibitemStop [0]{}%
\providecommand \bibitemNoStop [0]{.\EOS\space}%
\providecommand \EOS [0]{\spacefactor3000\relax}%
\providecommand \BibitemShut  [1]{\csname bibitem#1\endcsname}%
\let\auto@bib@innerbib\@empty
\bibitem [{\citenamefont {Harris}\ \emph {et~al.}(1967)\citenamefont {Harris},
  \citenamefont {Oshman},\ and\ \citenamefont {Byer}}]{Harris:67}%
  \BibitemOpen
  \bibfield  {author} {\bibinfo {author} {\bibfnamefont {S.~E.}\ \bibnamefont
  {Harris}}, \bibinfo {author} {\bibfnamefont {M.~K.}\ \bibnamefont {Oshman}},\
  and\ \bibinfo {author} {\bibfnamefont {R.~L.}\ \bibnamefont {Byer}},\
  }\bibfield  {title} {\bibinfo {title} {Observation of tunable optical
  parametric fluorescence},\ }\href
  {https://doi.org/10.1103/PhysRevLett.18.732} {\bibfield  {journal} {\bibinfo
  {journal} {Phys. Rev. Lett.}\ }\textbf {\bibinfo {volume} {18}},\ \bibinfo
  {pages} {732} (\bibinfo {year} {1967})}\BibitemShut {NoStop}%
\bibitem [{\citenamefont {Burnham}\ and\ \citenamefont
  {Weinberg}(1970)}]{Burnham1970}%
  \BibitemOpen
  \bibfield  {author} {\bibinfo {author} {\bibfnamefont {D.~C.}\ \bibnamefont
  {Burnham}}\ and\ \bibinfo {author} {\bibfnamefont {D.~L.}\ \bibnamefont
  {Weinberg}},\ }\bibfield  {title} {\bibinfo {title} {Observation of
  simultaneity in parametric production of optical photon pairs},\ }\href
  {https://doi.org/10.1103/PhysRevLett.25.84} {\bibfield  {journal} {\bibinfo
  {journal} {Phys. Rev. Lett.}\ }\textbf {\bibinfo {volume} {25}},\ \bibinfo
  {pages} {84} (\bibinfo {year} {1970})}\BibitemShut {NoStop}%
\bibitem [{\citenamefont {Takesue}\ and\ \citenamefont
  {Inoue}(2004)}]{Inoue2004}%
  \BibitemOpen
  \bibfield  {author} {\bibinfo {author} {\bibfnamefont {H.}~\bibnamefont
  {Takesue}}\ and\ \bibinfo {author} {\bibfnamefont {K.}~\bibnamefont
  {Inoue}},\ }\bibfield  {title} {\bibinfo {title} {Generation of
  polarization-entangled photon pairs and violation of bell's inequality using
  spontaneous four-wave mixing in a fiber loop},\ }\href
  {https://doi.org/10.1103/PhysRevA.70.031802} {\bibfield  {journal} {\bibinfo
  {journal} {Phys. Rev. A}\ }\textbf {\bibinfo {volume} {70}},\ \bibinfo
  {pages} {031802} (\bibinfo {year} {2004})}\BibitemShut {NoStop}%
\bibitem [{\citenamefont {Li}\ \emph {et~al.}(2004)\citenamefont {Li},
  \citenamefont {Chen}, \citenamefont {Voss}, \citenamefont {Sharping},\ and\
  \citenamefont {Kumar}}]{Li:04}%
  \BibitemOpen
  \bibfield  {author} {\bibinfo {author} {\bibfnamefont {X.}~\bibnamefont
  {Li}}, \bibinfo {author} {\bibfnamefont {J.}~\bibnamefont {Chen}}, \bibinfo
  {author} {\bibfnamefont {P.}~\bibnamefont {Voss}}, \bibinfo {author}
  {\bibfnamefont {J.}~\bibnamefont {Sharping}},\ and\ \bibinfo {author}
  {\bibfnamefont {P.}~\bibnamefont {Kumar}},\ }\bibfield  {title} {\bibinfo
  {title} {All-fiber photon-pair source for quantum communications: Improved
  generation of correlated photons},\ }\href
  {https://doi.org/10.1364/OPEX.12.003737} {\bibfield  {journal} {\bibinfo
  {journal} {Opt. Express}\ }\textbf {\bibinfo {volume} {12}},\ \bibinfo
  {pages} {3737} (\bibinfo {year} {2004})}\BibitemShut {NoStop}%
\bibitem [{\citenamefont {Clemmen}\ \emph {et~al.}(2009)\citenamefont
  {Clemmen}, \citenamefont {Huy}, \citenamefont {Bogaerts}, \citenamefont
  {Baets}, \citenamefont {Emplit},\ and\ \citenamefont
  {Massar}}]{clemmen2009continuous}%
  \BibitemOpen
  \bibfield  {author} {\bibinfo {author} {\bibfnamefont {S.}~\bibnamefont
  {Clemmen}}, \bibinfo {author} {\bibfnamefont {K.~P.}\ \bibnamefont {Huy}},
  \bibinfo {author} {\bibfnamefont {W.}~\bibnamefont {Bogaerts}}, \bibinfo
  {author} {\bibfnamefont {R.~G.}\ \bibnamefont {Baets}}, \bibinfo {author}
  {\bibfnamefont {P.}~\bibnamefont {Emplit}},\ and\ \bibinfo {author}
  {\bibfnamefont {S.}~\bibnamefont {Massar}},\ }\bibfield  {title} {\bibinfo
  {title} {Continuous wave photon pair generation in silicon-on-insulator
  waveguides and ring resonators},\ }\href@noop {} {\bibfield  {journal}
  {\bibinfo  {journal} {Optics express}\ }\textbf {\bibinfo {volume} {17}},\
  \bibinfo {pages} {16558} (\bibinfo {year} {2009})}\BibitemShut {NoStop}%
\bibitem [{\citenamefont {Caspani}\ \emph {et~al.}(2017)\citenamefont
  {Caspani}, \citenamefont {Xiong}, \citenamefont {Eggleton}, \citenamefont
  {Bajoni}, \citenamefont {Liscidini}, \citenamefont {Galli}, \citenamefont
  {Morandotti},\ and\ \citenamefont {Moss}}]{caspani2017integrated}%
  \BibitemOpen
  \bibfield  {author} {\bibinfo {author} {\bibfnamefont {L.}~\bibnamefont
  {Caspani}}, \bibinfo {author} {\bibfnamefont {C.}~\bibnamefont {Xiong}},
  \bibinfo {author} {\bibfnamefont {B.~J.}\ \bibnamefont {Eggleton}}, \bibinfo
  {author} {\bibfnamefont {D.}~\bibnamefont {Bajoni}}, \bibinfo {author}
  {\bibfnamefont {M.}~\bibnamefont {Liscidini}}, \bibinfo {author}
  {\bibfnamefont {M.}~\bibnamefont {Galli}}, \bibinfo {author} {\bibfnamefont
  {R.}~\bibnamefont {Morandotti}},\ and\ \bibinfo {author} {\bibfnamefont
  {D.~J.}\ \bibnamefont {Moss}},\ }\bibfield  {title} {\bibinfo {title}
  {Integrated sources of photon quantum states based on nonlinear optics},\
  }\href@noop {} {\bibfield  {journal} {\bibinfo  {journal} {Light: Science \&
  Applications}\ }\textbf {\bibinfo {volume} {6}},\ \bibinfo {pages} {e17100}
  (\bibinfo {year} {2017})}\BibitemShut {NoStop}%
\bibitem [{\citenamefont {Azzini}\ \emph {et~al.}(2012)\citenamefont {Azzini},
  \citenamefont {Grassani}, \citenamefont {Galli}, \citenamefont {Andreani},
  \citenamefont {Sorel}, \citenamefont {Strain}, \citenamefont {Helt},
  \citenamefont {Sipe}, \citenamefont {Liscidini},\ and\ \citenamefont
  {Bajoni}}]{azzini2012classical}%
  \BibitemOpen
  \bibfield  {author} {\bibinfo {author} {\bibfnamefont {S.}~\bibnamefont
  {Azzini}}, \bibinfo {author} {\bibfnamefont {D.}~\bibnamefont {Grassani}},
  \bibinfo {author} {\bibfnamefont {M.}~\bibnamefont {Galli}}, \bibinfo
  {author} {\bibfnamefont {L.~C.}\ \bibnamefont {Andreani}}, \bibinfo {author}
  {\bibfnamefont {M.}~\bibnamefont {Sorel}}, \bibinfo {author} {\bibfnamefont
  {M.~J.}\ \bibnamefont {Strain}}, \bibinfo {author} {\bibfnamefont
  {L.}~\bibnamefont {Helt}}, \bibinfo {author} {\bibfnamefont {J.}~\bibnamefont
  {Sipe}}, \bibinfo {author} {\bibfnamefont {M.}~\bibnamefont {Liscidini}},\
  and\ \bibinfo {author} {\bibfnamefont {D.}~\bibnamefont {Bajoni}},\
  }\bibfield  {title} {\bibinfo {title} {From classical four-wave mixing to
  parametric fluorescence in silicon microring resonators},\ }\href@noop {}
  {\bibfield  {journal} {\bibinfo  {journal} {Optics letters}\ }\textbf
  {\bibinfo {volume} {37}},\ \bibinfo {pages} {3807} (\bibinfo {year}
  {2012})}\BibitemShut {NoStop}%
\bibitem [{\citenamefont {Imany}\ \emph {et~al.}(2018)\citenamefont {Imany},
  \citenamefont {Jaramillo-Villegas}, \citenamefont {Odele}, \citenamefont
  {Han}, \citenamefont {Leaird}, \citenamefont {Lukens}, \citenamefont
  {Lougovski}, \citenamefont {Qi},\ and\ \citenamefont {Weiner}}]{Imany:18}%
  \BibitemOpen
  \bibfield  {author} {\bibinfo {author} {\bibfnamefont {P.}~\bibnamefont
  {Imany}}, \bibinfo {author} {\bibfnamefont {J.~A.}\ \bibnamefont
  {Jaramillo-Villegas}}, \bibinfo {author} {\bibfnamefont {O.~D.}\ \bibnamefont
  {Odele}}, \bibinfo {author} {\bibfnamefont {K.}~\bibnamefont {Han}}, \bibinfo
  {author} {\bibfnamefont {D.~E.}\ \bibnamefont {Leaird}}, \bibinfo {author}
  {\bibfnamefont {J.~M.}\ \bibnamefont {Lukens}}, \bibinfo {author}
  {\bibfnamefont {P.}~\bibnamefont {Lougovski}}, \bibinfo {author}
  {\bibfnamefont {M.}~\bibnamefont {Qi}},\ and\ \bibinfo {author}
  {\bibfnamefont {A.~M.}\ \bibnamefont {Weiner}},\ }\bibfield  {title}
  {\bibinfo {title} {50-ghz-spaced comb of high-dimensional frequency-bin
  entangled photons from an on-chip silicon nitride microresonator},\ }\href
  {https://doi.org/10.1364/OE.26.001825} {\bibfield  {journal} {\bibinfo
  {journal} {Opt. Express}\ }\textbf {\bibinfo {volume} {26}},\ \bibinfo
  {pages} {1825} (\bibinfo {year} {2018})}\BibitemShut {NoStop}%
\bibitem [{\citenamefont {Kues}\ \emph {et~al.}(2017)\citenamefont {Kues},
  \citenamefont {Reimer}, \citenamefont {Roztocki}, \citenamefont {Cort{\'e}s},
  \citenamefont {Sciara}, \citenamefont {Wetzel}, \citenamefont {Zhang},
  \citenamefont {Cino}, \citenamefont {Chu}, \citenamefont {Little},
  \citenamefont {Moss}, \citenamefont {Caspani}, \citenamefont {Azaña},\ and\
  \citenamefont {Morandotti}}]{kues2017chip}%
  \BibitemOpen
  \bibfield  {author} {\bibinfo {author} {\bibfnamefont {M.}~\bibnamefont
  {Kues}}, \bibinfo {author} {\bibfnamefont {C.}~\bibnamefont {Reimer}},
  \bibinfo {author} {\bibfnamefont {P.}~\bibnamefont {Roztocki}}, \bibinfo
  {author} {\bibfnamefont {L.~R.}\ \bibnamefont {Cort{\'e}s}}, \bibinfo
  {author} {\bibfnamefont {S.}~\bibnamefont {Sciara}}, \bibinfo {author}
  {\bibfnamefont {B.}~\bibnamefont {Wetzel}}, \bibinfo {author} {\bibfnamefont
  {Y.}~\bibnamefont {Zhang}}, \bibinfo {author} {\bibfnamefont
  {A.}~\bibnamefont {Cino}}, \bibinfo {author} {\bibfnamefont {S.~T.}\
  \bibnamefont {Chu}}, \bibinfo {author} {\bibfnamefont {B.~E.}\ \bibnamefont
  {Little}}, \bibinfo {author} {\bibfnamefont {D.~J.~M.}\ \bibnamefont {Moss}},
  \bibinfo {author} {\bibfnamefont {L.}~\bibnamefont {Caspani}}, \bibinfo
  {author} {\bibfnamefont {J.}~\bibnamefont {Azaña}},\ and\ \bibinfo {author}
  {\bibfnamefont {R.}~\bibnamefont {Morandotti}},\ }\bibfield  {title}
  {\bibinfo {title} {On-chip generation of high-dimensional entangled quantum
  states and their coherent control},\ }\href@noop {} {\bibfield  {journal}
  {\bibinfo  {journal} {Nature}\ }\textbf {\bibinfo {volume} {546}},\ \bibinfo
  {pages} {622} (\bibinfo {year} {2017})}\BibitemShut {NoStop}%
\bibitem [{\citenamefont {Orieux}\ \emph {et~al.}(2013)\citenamefont {Orieux},
  \citenamefont {Eckstein}, \citenamefont {Lema\^{\i}tre}, \citenamefont
  {Filloux}, \citenamefont {Favero}, \citenamefont {Leo}, \citenamefont
  {Coudreau}, \citenamefont {Keller}, \citenamefont {Milman},\ and\
  \citenamefont {Ducci}}]{ducci2013}%
  \BibitemOpen
  \bibfield  {author} {\bibinfo {author} {\bibfnamefont {A.}~\bibnamefont
  {Orieux}}, \bibinfo {author} {\bibfnamefont {A.}~\bibnamefont {Eckstein}},
  \bibinfo {author} {\bibfnamefont {A.}~\bibnamefont {Lema\^{\i}tre}}, \bibinfo
  {author} {\bibfnamefont {P.}~\bibnamefont {Filloux}}, \bibinfo {author}
  {\bibfnamefont {I.}~\bibnamefont {Favero}}, \bibinfo {author} {\bibfnamefont
  {G.}~\bibnamefont {Leo}}, \bibinfo {author} {\bibfnamefont {T.}~\bibnamefont
  {Coudreau}}, \bibinfo {author} {\bibfnamefont {A.}~\bibnamefont {Keller}},
  \bibinfo {author} {\bibfnamefont {P.}~\bibnamefont {Milman}},\ and\ \bibinfo
  {author} {\bibfnamefont {S.}~\bibnamefont {Ducci}},\ }\bibfield  {title}
  {\bibinfo {title} {Direct bell states generation on a iii-v semiconductor
  chip at room temperature},\ }\href
  {https://doi.org/10.1103/PhysRevLett.110.160502} {\bibfield  {journal}
  {\bibinfo  {journal} {Phys. Rev. Lett.}\ }\textbf {\bibinfo {volume} {110}},\
  \bibinfo {pages} {160502} (\bibinfo {year} {2013})}\BibitemShut {NoStop}%
\bibitem [{\citenamefont {Sharping}\ \emph {et~al.}(2006)\citenamefont
  {Sharping}, \citenamefont {Lee}, \citenamefont {Foster}, \citenamefont
  {Turner}, \citenamefont {Schmidt}, \citenamefont {Lipson}, \citenamefont
  {Gaeta},\ and\ \citenamefont {Kumar}}]{sharping2006generation}%
  \BibitemOpen
  \bibfield  {author} {\bibinfo {author} {\bibfnamefont {J.~E.}\ \bibnamefont
  {Sharping}}, \bibinfo {author} {\bibfnamefont {K.~F.}\ \bibnamefont {Lee}},
  \bibinfo {author} {\bibfnamefont {M.~A.}\ \bibnamefont {Foster}}, \bibinfo
  {author} {\bibfnamefont {A.~C.}\ \bibnamefont {Turner}}, \bibinfo {author}
  {\bibfnamefont {B.~S.}\ \bibnamefont {Schmidt}}, \bibinfo {author}
  {\bibfnamefont {M.}~\bibnamefont {Lipson}}, \bibinfo {author} {\bibfnamefont
  {A.~L.}\ \bibnamefont {Gaeta}},\ and\ \bibinfo {author} {\bibfnamefont
  {P.}~\bibnamefont {Kumar}},\ }\bibfield  {title} {\bibinfo {title}
  {Generation of correlated photons in nanoscale silicon waveguides},\
  }\href@noop {} {\bibfield  {journal} {\bibinfo  {journal} {Optics express}\
  }\textbf {\bibinfo {volume} {14}},\ \bibinfo {pages} {12388} (\bibinfo {year}
  {2006})}\BibitemShut {NoStop}%
\bibitem [{\citenamefont {Takesue}\ \emph {et~al.}(2007)\citenamefont
  {Takesue}, \citenamefont {Tokura}, \citenamefont {Fukuda}, \citenamefont
  {Tsuchizawa}, \citenamefont {Watanabe}, \citenamefont {Yamada},\ and\
  \citenamefont {Itabashi}}]{takesue2007entanglement}%
  \BibitemOpen
  \bibfield  {author} {\bibinfo {author} {\bibfnamefont {H.}~\bibnamefont
  {Takesue}}, \bibinfo {author} {\bibfnamefont {Y.}~\bibnamefont {Tokura}},
  \bibinfo {author} {\bibfnamefont {H.}~\bibnamefont {Fukuda}}, \bibinfo
  {author} {\bibfnamefont {T.}~\bibnamefont {Tsuchizawa}}, \bibinfo {author}
  {\bibfnamefont {T.}~\bibnamefont {Watanabe}}, \bibinfo {author}
  {\bibfnamefont {K.}~\bibnamefont {Yamada}},\ and\ \bibinfo {author}
  {\bibfnamefont {S.-i.}\ \bibnamefont {Itabashi}},\ }\bibfield  {title}
  {\bibinfo {title} {Entanglement generation using silicon wire waveguide},\
  }\href@noop {} {\bibfield  {journal} {\bibinfo  {journal} {Applied Physics
  Letters}\ }\textbf {\bibinfo {volume} {91}},\ \bibinfo {pages} {201108}
  (\bibinfo {year} {2007})}\BibitemShut {NoStop}%
\bibitem [{\citenamefont {Gentry}\ \emph {et~al.}(2014)\citenamefont {Gentry},
  \citenamefont {Zeng},\ and\ \citenamefont {Popovi\'{c}}}]{Gentry2014}%
  \BibitemOpen
  \bibfield  {author} {\bibinfo {author} {\bibfnamefont {C.~M.}\ \bibnamefont
  {Gentry}}, \bibinfo {author} {\bibfnamefont {X.}~\bibnamefont {Zeng}},\ and\
  \bibinfo {author} {\bibfnamefont {M.~A.}\ \bibnamefont {Popovi\'{c}}},\
  }\bibfield  {title} {\bibinfo {title} {Tunable coupled-mode dispersion
  compensation and its application to on-chip resonant four-wave mixing},\
  }\href {https://doi.org/10.1364/OL.39.005689} {\bibfield  {journal} {\bibinfo
   {journal} {Opt. Lett.}\ }\textbf {\bibinfo {volume} {39}},\ \bibinfo {pages}
  {5689} (\bibinfo {year} {2014})}\BibitemShut {NoStop}%
\bibitem [{\citenamefont {Zeng}\ \emph {et~al.}(2015)\citenamefont {Zeng},
  \citenamefont {Gentry},\ and\ \citenamefont {Popovi\'{c}}}]{Zeng2015}%
  \BibitemOpen
  \bibfield  {author} {\bibinfo {author} {\bibfnamefont {X.}~\bibnamefont
  {Zeng}}, \bibinfo {author} {\bibfnamefont {C.~M.}\ \bibnamefont {Gentry}},\
  and\ \bibinfo {author} {\bibfnamefont {M.~A.}\ \bibnamefont {Popovi\'{c}}},\
  }\bibfield  {title} {\bibinfo {title} {Four-wave mixing in silicon
  coupled-cavity resonators with port-selective, orthogonal supermode
  excitation},\ }\href {https://doi.org/10.1364/OL.40.002120} {\bibfield
  {journal} {\bibinfo  {journal} {Opt. Lett.}\ }\textbf {\bibinfo {volume}
  {40}},\ \bibinfo {pages} {2120} (\bibinfo {year} {2015})}\BibitemShut
  {NoStop}%
\bibitem [{\citenamefont {Davanco}\ \emph {et~al.}(2012)\citenamefont
  {Davanco}, \citenamefont {Ong}, \citenamefont {Shehata}, \citenamefont
  {Tosi}, \citenamefont {Agha}, \citenamefont {Assefa}, \citenamefont {Xia},
  \citenamefont {Green}, \citenamefont {Mookherjea},\ and\ \citenamefont
  {Srinivasan}}]{davanco2012telecommunications}%
  \BibitemOpen
  \bibfield  {author} {\bibinfo {author} {\bibfnamefont {M.}~\bibnamefont
  {Davanco}}, \bibinfo {author} {\bibfnamefont {J.~R.}\ \bibnamefont {Ong}},
  \bibinfo {author} {\bibfnamefont {A.~B.}\ \bibnamefont {Shehata}}, \bibinfo
  {author} {\bibfnamefont {A.}~\bibnamefont {Tosi}}, \bibinfo {author}
  {\bibfnamefont {I.}~\bibnamefont {Agha}}, \bibinfo {author} {\bibfnamefont
  {S.}~\bibnamefont {Assefa}}, \bibinfo {author} {\bibfnamefont
  {F.}~\bibnamefont {Xia}}, \bibinfo {author} {\bibfnamefont {W.~M.}\
  \bibnamefont {Green}}, \bibinfo {author} {\bibfnamefont {S.}~\bibnamefont
  {Mookherjea}},\ and\ \bibinfo {author} {\bibfnamefont {K.}~\bibnamefont
  {Srinivasan}},\ }\bibfield  {title} {\bibinfo {title}
  {Telecommunications-band heralded single photons from a silicon nanophotonic
  chip},\ }\href@noop {} {\bibfield  {journal} {\bibinfo  {journal} {Applied
  Physics Letters}\ }\textbf {\bibinfo {volume} {100}},\ \bibinfo {pages}
  {261104} (\bibinfo {year} {2012})}\BibitemShut {NoStop}%
\bibitem [{\citenamefont {Xiong}\ \emph {et~al.}(2011)\citenamefont {Xiong},
  \citenamefont {Monat}, \citenamefont {Clark}, \citenamefont {Grillet},
  \citenamefont {Marshall}, \citenamefont {Steel}, \citenamefont {Li},
  \citenamefont {O’Faolain}, \citenamefont {Krauss}, \citenamefont {Rarity},\
  and\ \citenamefont {Eggleton}}]{xiong2011slow}%
  \BibitemOpen
  \bibfield  {author} {\bibinfo {author} {\bibfnamefont {C.}~\bibnamefont
  {Xiong}}, \bibinfo {author} {\bibfnamefont {C.}~\bibnamefont {Monat}},
  \bibinfo {author} {\bibfnamefont {A.~S.}\ \bibnamefont {Clark}}, \bibinfo
  {author} {\bibfnamefont {C.}~\bibnamefont {Grillet}}, \bibinfo {author}
  {\bibfnamefont {G.~D.}\ \bibnamefont {Marshall}}, \bibinfo {author}
  {\bibfnamefont {M.~J.}\ \bibnamefont {Steel}}, \bibinfo {author}
  {\bibfnamefont {J.}~\bibnamefont {Li}}, \bibinfo {author} {\bibfnamefont
  {L.}~\bibnamefont {O’Faolain}}, \bibinfo {author} {\bibfnamefont {T.~F.}\
  \bibnamefont {Krauss}}, \bibinfo {author} {\bibfnamefont {J.~G.}\
  \bibnamefont {Rarity}},\ and\ \bibinfo {author} {\bibfnamefont {B.~J.}\
  \bibnamefont {Eggleton}},\ }\bibfield  {title} {\bibinfo {title} {Slow-light
  enhanced correlated photon pair generation in a silicon photonic crystal
  waveguide},\ }\href@noop {} {\bibfield  {journal} {\bibinfo  {journal}
  {Optics letters}\ }\textbf {\bibinfo {volume} {36}},\ \bibinfo {pages} {3413}
  (\bibinfo {year} {2011})}\BibitemShut {NoStop}%
\bibitem [{\citenamefont {Mittal}\ \emph {et~al.}(2018)\citenamefont {Mittal},
  \citenamefont {Goldschmidt},\ and\ \citenamefont {Hafezi}}]{Mittal2018}%
  \BibitemOpen
  \bibfield  {author} {\bibinfo {author} {\bibfnamefont {S.}~\bibnamefont
  {Mittal}}, \bibinfo {author} {\bibfnamefont {E.~A.}\ \bibnamefont
  {Goldschmidt}},\ and\ \bibinfo {author} {\bibfnamefont {M.}~\bibnamefont
  {Hafezi}},\ }\bibfield  {title} {\bibinfo {title} {A topological source of
  quantum light},\ }\href {https://doi.org/10.1038/s41586-018-0478-3}
  {\bibfield  {journal} {\bibinfo  {journal} {Nature}\ }\textbf {\bibinfo
  {volume} {561}},\ \bibinfo {pages} {502} (\bibinfo {year}
  {2018})}\BibitemShut {NoStop}%
\bibitem [{\citenamefont {Blanco-Redondo}\ \emph {et~al.}(2018)\citenamefont
  {Blanco-Redondo}, \citenamefont {Bell}, \citenamefont {Oren}, \citenamefont
  {Eggleton},\ and\ \citenamefont {Segev}}]{Blanco2018}%
  \BibitemOpen
  \bibfield  {author} {\bibinfo {author} {\bibfnamefont {A.}~\bibnamefont
  {Blanco-Redondo}}, \bibinfo {author} {\bibfnamefont {B.}~\bibnamefont
  {Bell}}, \bibinfo {author} {\bibfnamefont {D.}~\bibnamefont {Oren}}, \bibinfo
  {author} {\bibfnamefont {B.~J.}\ \bibnamefont {Eggleton}},\ and\ \bibinfo
  {author} {\bibfnamefont {M.}~\bibnamefont {Segev}},\ }\bibfield  {title}
  {\bibinfo {title} {Topological protection of biphoton states},\ }\href
  {https://doi.org/10.1126/science.aau4296} {\bibfield  {journal} {\bibinfo
  {journal} {Science}\ }\textbf {\bibinfo {volume} {362}},\ \bibinfo {pages}
  {568} (\bibinfo {year} {2018})}\BibitemShut {NoStop}%
\bibitem [{\citenamefont {Zhang}\ \emph {et~al.}(2021)\citenamefont {Zhang},
  \citenamefont {Menotti}, \citenamefont {Tan}, \citenamefont {Vaidya},
  \citenamefont {Mahler}, \citenamefont {Helt}, \citenamefont {Zatti},
  \citenamefont {Liscidini}, \citenamefont {Morrison},\ and\ \citenamefont
  {Vernon}}]{zhang2021}%
  \BibitemOpen
  \bibfield  {author} {\bibinfo {author} {\bibfnamefont {Y.}~\bibnamefont
  {Zhang}}, \bibinfo {author} {\bibfnamefont {M.}~\bibnamefont {Menotti}},
  \bibinfo {author} {\bibfnamefont {K.}~\bibnamefont {Tan}}, \bibinfo {author}
  {\bibfnamefont {V.~D.}\ \bibnamefont {Vaidya}}, \bibinfo {author}
  {\bibfnamefont {D.~H.}\ \bibnamefont {Mahler}}, \bibinfo {author}
  {\bibfnamefont {L.~G.}\ \bibnamefont {Helt}}, \bibinfo {author}
  {\bibfnamefont {L.}~\bibnamefont {Zatti}}, \bibinfo {author} {\bibfnamefont
  {M.}~\bibnamefont {Liscidini}}, \bibinfo {author} {\bibfnamefont
  {B.}~\bibnamefont {Morrison}},\ and\ \bibinfo {author} {\bibfnamefont
  {Z.}~\bibnamefont {Vernon}},\ }\bibfield  {title} {\bibinfo {title} {Squeezed
  light from a nanophotonic molecule},\ }\href
  {https://doi.org/10.1038/s41467-021-22540-2} {\bibfield  {journal} {\bibinfo
  {journal} {Nature Communications}\ }\textbf {\bibinfo {volume} {12}},\
  \bibinfo {pages} {2233} (\bibinfo {year} {2021})}\BibitemShut {NoStop}%
\bibitem [{\citenamefont {Menotti}\ \emph {et~al.}(2019)\citenamefont
  {Menotti}, \citenamefont {Morrison}, \citenamefont {Tan}, \citenamefont
  {Vernon}, \citenamefont {Sipe},\ and\ \citenamefont
  {Liscidini}}]{menotti2019nonlinear}%
  \BibitemOpen
  \bibfield  {author} {\bibinfo {author} {\bibfnamefont {M.}~\bibnamefont
  {Menotti}}, \bibinfo {author} {\bibfnamefont {B.}~\bibnamefont {Morrison}},
  \bibinfo {author} {\bibfnamefont {K.}~\bibnamefont {Tan}}, \bibinfo {author}
  {\bibfnamefont {Z.}~\bibnamefont {Vernon}}, \bibinfo {author} {\bibfnamefont
  {J.}~\bibnamefont {Sipe}},\ and\ \bibinfo {author} {\bibfnamefont
  {M.}~\bibnamefont {Liscidini}},\ }\bibfield  {title} {\bibinfo {title}
  {Nonlinear coupling of linearly uncoupled resonators},\ }\href@noop {}
  {\bibfield  {journal} {\bibinfo  {journal} {Physical review letters}\
  }\textbf {\bibinfo {volume} {122}},\ \bibinfo {pages} {013904} (\bibinfo
  {year} {2019})}\BibitemShut {NoStop}%
\bibitem [{\citenamefont {Tan}\ \emph {et~al.}(2020)\citenamefont {Tan},
  \citenamefont {Menotti}, \citenamefont {Vernon}, \citenamefont {Sipe},
  \citenamefont {Liscidini},\ and\ \citenamefont {Morrison}}]{Tan:20}%
  \BibitemOpen
  \bibfield  {author} {\bibinfo {author} {\bibfnamefont {K.}~\bibnamefont
  {Tan}}, \bibinfo {author} {\bibfnamefont {M.}~\bibnamefont {Menotti}},
  \bibinfo {author} {\bibfnamefont {Z.}~\bibnamefont {Vernon}}, \bibinfo
  {author} {\bibfnamefont {J.~E.}\ \bibnamefont {Sipe}}, \bibinfo {author}
  {\bibfnamefont {M.}~\bibnamefont {Liscidini}},\ and\ \bibinfo {author}
  {\bibfnamefont {B.}~\bibnamefont {Morrison}},\ }\bibfield  {title} {\bibinfo
  {title} {Stimulated four-wave mixing in linearly uncoupled resonators},\
  }\href {https://doi.org/10.1364/OL.381563} {\bibfield  {journal} {\bibinfo
  {journal} {Opt. Lett.}\ }\textbf {\bibinfo {volume} {45}},\ \bibinfo {pages}
  {873} (\bibinfo {year} {2020})}\BibitemShut {NoStop}%
\bibitem [{\citenamefont {Starling}\ \emph {et~al.}(2020)\citenamefont
  {Starling}, \citenamefont {Poirier}, \citenamefont {Fanto}, \citenamefont
  {Steidle}, \citenamefont {Tison}, \citenamefont {Howland},\ and\
  \citenamefont {Preble}}]{Preble20}%
  \BibitemOpen
  \bibfield  {author} {\bibinfo {author} {\bibfnamefont {D.~J.}\ \bibnamefont
  {Starling}}, \bibinfo {author} {\bibfnamefont {J.}~\bibnamefont {Poirier}},
  \bibinfo {author} {\bibfnamefont {M.}~\bibnamefont {Fanto}}, \bibinfo
  {author} {\bibfnamefont {J.~A.}\ \bibnamefont {Steidle}}, \bibinfo {author}
  {\bibfnamefont {C.~C.}\ \bibnamefont {Tison}}, \bibinfo {author}
  {\bibfnamefont {G.~A.}\ \bibnamefont {Howland}},\ and\ \bibinfo {author}
  {\bibfnamefont {S.~F.}\ \bibnamefont {Preble}},\ }\bibfield  {title}
  {\bibinfo {title} {Nonlinear photon pair generation in a highly dispersive
  medium},\ }\href {https://doi.org/10.1103/PhysRevApplied.13.041005}
  {\bibfield  {journal} {\bibinfo  {journal} {Phys. Rev. Applied}\ }\textbf
  {\bibinfo {volume} {13}},\ \bibinfo {pages} {041005} (\bibinfo {year}
  {2020})}\BibitemShut {NoStop}%
\bibitem [{\citenamefont {Sabattoli}\ \emph {et~al.}(2021)\citenamefont
  {Sabattoli}, \citenamefont {El~Dirani}, \citenamefont {Youssef},
  \citenamefont {Garrisi}, \citenamefont {Grassani}, \citenamefont {Zatti},
  \citenamefont {Petit-Etienne}, \citenamefont {Pargon}, \citenamefont {Sipe},
  \citenamefont {Liscidini}, \citenamefont {Sciancalepore}, \citenamefont
  {Bajoni},\ and\ \citenamefont {Galli}}]{sabattoli2021}%
  \BibitemOpen
  \bibfield  {author} {\bibinfo {author} {\bibfnamefont {F.~A.}\ \bibnamefont
  {Sabattoli}}, \bibinfo {author} {\bibfnamefont {H.}~\bibnamefont
  {El~Dirani}}, \bibinfo {author} {\bibfnamefont {L.}~\bibnamefont {Youssef}},
  \bibinfo {author} {\bibfnamefont {F.}~\bibnamefont {Garrisi}}, \bibinfo
  {author} {\bibfnamefont {D.}~\bibnamefont {Grassani}}, \bibinfo {author}
  {\bibfnamefont {L.}~\bibnamefont {Zatti}}, \bibinfo {author} {\bibfnamefont
  {C.}~\bibnamefont {Petit-Etienne}}, \bibinfo {author} {\bibfnamefont
  {E.}~\bibnamefont {Pargon}}, \bibinfo {author} {\bibfnamefont {J.~E.}\
  \bibnamefont {Sipe}}, \bibinfo {author} {\bibfnamefont {M.}~\bibnamefont
  {Liscidini}}, \bibinfo {author} {\bibfnamefont {C.}~\bibnamefont
  {Sciancalepore}}, \bibinfo {author} {\bibfnamefont {D.}~\bibnamefont
  {Bajoni}},\ and\ \bibinfo {author} {\bibfnamefont {M.}~\bibnamefont
  {Galli}},\ }\bibfield  {title} {\bibinfo {title} {Suppression of parasitic
  nonlinear processes in spontaneous four-wave mixing with linearly uncoupled
  resonators},\ }\href {https://doi.org/10.1103/PhysRevLett.127.033901}
  {\bibfield  {journal} {\bibinfo  {journal} {Phys. Rev. Lett.}\ }\textbf
  {\bibinfo {volume} {127}},\ \bibinfo {pages} {033901} (\bibinfo {year}
  {2021})}\BibitemShut {NoStop}%
\bibitem [{\citenamefont {Yariv}\ and\ \citenamefont {Yeh}(2007)}]{photYY}%
  \BibitemOpen
  \bibfield  {author} {\bibinfo {author} {\bibfnamefont {A.}~\bibnamefont
  {Yariv}}\ and\ \bibinfo {author} {\bibfnamefont {P.}~\bibnamefont {Yeh}},\
  }\href@noop {} {\emph {\bibinfo {title} {Photonics}}}\ (\bibinfo  {publisher}
  {Oxford University Press},\ \bibinfo {year} {2007})\BibitemShut {NoStop}%
\bibitem [{\citenamefont {Heebner}\ \emph {et~al.}(2008)\citenamefont
  {Heebner}, \citenamefont {Grover},\ and\ \citenamefont
  {Ibrahim}}]{heebner2008optical}%
  \BibitemOpen
  \bibfield  {author} {\bibinfo {author} {\bibfnamefont {J.}~\bibnamefont
  {Heebner}}, \bibinfo {author} {\bibfnamefont {R.}~\bibnamefont {Grover}},\
  and\ \bibinfo {author} {\bibfnamefont {T.}~\bibnamefont {Ibrahim}},\
  }\href@noop {} {\emph {\bibinfo {title} {Optical microresonators}}}\
  (\bibinfo  {publisher} {Springer New York, NY},\ \bibinfo {year}
  {2008})\BibitemShut {NoStop}%
\bibitem [{\citenamefont {Banic}\ \emph {et~al.}(2022)\citenamefont {Banic},
  \citenamefont {Zatti}, \citenamefont {Liscidini},\ and\ \citenamefont
  {Sipe}}]{banic2022two}%
  \BibitemOpen
  \bibfield  {author} {\bibinfo {author} {\bibfnamefont {M.}~\bibnamefont
  {Banic}}, \bibinfo {author} {\bibfnamefont {L.}~\bibnamefont {Zatti}},
  \bibinfo {author} {\bibfnamefont {M.}~\bibnamefont {Liscidini}},\ and\
  \bibinfo {author} {\bibfnamefont {J.}~\bibnamefont {Sipe}},\ }\bibfield
  {title} {\bibinfo {title} {Two strategies for modeling nonlinear optics in
  lossy integrated photonic structures},\ }\href@noop {} {\bibfield  {journal}
  {\bibinfo  {journal} {Physical Review A}\ }\textbf {\bibinfo {volume}
  {106}},\ \bibinfo {pages} {043707} (\bibinfo {year} {2022})}\BibitemShut
  {NoStop}%
\bibitem [{\citenamefont {Yang}\ \emph {et~al.}(2008)\citenamefont {Yang},
  \citenamefont {Liscidini},\ and\ \citenamefont {Sipe}}]{yang2008spontaneous}%
  \BibitemOpen
  \bibfield  {author} {\bibinfo {author} {\bibfnamefont {Z.}~\bibnamefont
  {Yang}}, \bibinfo {author} {\bibfnamefont {M.}~\bibnamefont {Liscidini}},\
  and\ \bibinfo {author} {\bibfnamefont {J.}~\bibnamefont {Sipe}},\ }\bibfield
  {title} {\bibinfo {title} {Spontaneous parametric down-conversion in
  waveguides: a backward heisenberg picture approach},\ }\href@noop {}
  {\bibfield  {journal} {\bibinfo  {journal} {Physical Review A}\ }\textbf
  {\bibinfo {volume} {77}},\ \bibinfo {pages} {033808} (\bibinfo {year}
  {2008})}\BibitemShut {NoStop}%
\bibitem [{\citenamefont {Sabattoli}\ \emph {et~al.}(2022)\citenamefont
  {Sabattoli}, \citenamefont {Dirani}, \citenamefont {Youssef}, \citenamefont
  {Gianini}, \citenamefont {Zatti}, \citenamefont {Garrisi}, \citenamefont
  {Grassani}, \citenamefont {Petit-Etienne}, \citenamefont {Pargon},
  \citenamefont {Sipe}, \citenamefont {Liscidini}, \citenamefont
  {Sciancalepore}, \citenamefont {Bajoni},\ and\ \citenamefont
  {Galli}}]{sabattoli2022nonlinear}%
  \BibitemOpen
  \bibfield  {author} {\bibinfo {author} {\bibfnamefont {F.}~\bibnamefont
  {Sabattoli}}, \bibinfo {author} {\bibfnamefont {H.~E.}\ \bibnamefont
  {Dirani}}, \bibinfo {author} {\bibfnamefont {L.}~\bibnamefont {Youssef}},
  \bibinfo {author} {\bibfnamefont {L.}~\bibnamefont {Gianini}}, \bibinfo
  {author} {\bibfnamefont {L.}~\bibnamefont {Zatti}}, \bibinfo {author}
  {\bibfnamefont {F.}~\bibnamefont {Garrisi}}, \bibinfo {author} {\bibfnamefont
  {D.}~\bibnamefont {Grassani}}, \bibinfo {author} {\bibfnamefont
  {C.}~\bibnamefont {Petit-Etienne}}, \bibinfo {author} {\bibfnamefont
  {E.}~\bibnamefont {Pargon}}, \bibinfo {author} {\bibfnamefont
  {J.}~\bibnamefont {Sipe}}, \bibinfo {author} {\bibfnamefont {M.}~\bibnamefont
  {Liscidini}}, \bibinfo {author} {\bibfnamefont {C.}~\bibnamefont
  {Sciancalepore}}, \bibinfo {author} {\bibfnamefont {D.}~\bibnamefont
  {Bajoni}},\ and\ \bibinfo {author} {\bibfnamefont {M.}~\bibnamefont
  {Galli}},\ }\bibfield  {title} {\bibinfo {title} {Nonlinear coupling of
  linearly uncoupled resonators through a mach-zehnder interferometer},\
  }\href@noop {} {\bibfield  {journal} {\bibinfo  {journal} {Applied Physics
  Letters}\ }\textbf {\bibinfo {volume} {121}},\ \bibinfo {pages} {201101}
  (\bibinfo {year} {2022})}\BibitemShut {NoStop}%
\bibitem [{\citenamefont {Onodera}\ \emph {et~al.}(2016)\citenamefont
  {Onodera}, \citenamefont {Liscidini}, \citenamefont {Sipe},\ and\
  \citenamefont {Helt}}]{onodera2016parametric}%
  \BibitemOpen
  \bibfield  {author} {\bibinfo {author} {\bibfnamefont {T.}~\bibnamefont
  {Onodera}}, \bibinfo {author} {\bibfnamefont {M.}~\bibnamefont {Liscidini}},
  \bibinfo {author} {\bibfnamefont {J.}~\bibnamefont {Sipe}},\ and\ \bibinfo
  {author} {\bibfnamefont {L.}~\bibnamefont {Helt}},\ }\bibfield  {title}
  {\bibinfo {title} {Parametric fluorescence in a sequence of resonators: An
  analogy with dicke superradiance},\ }\href@noop {} {\bibfield  {journal}
  {\bibinfo  {journal} {Physical Review A}\ }\textbf {\bibinfo {volume} {93}},\
  \bibinfo {pages} {043837} (\bibinfo {year} {2016})}\BibitemShut {NoStop}%
\bibitem [{\citenamefont {Helt}\ \emph {et~al.}(2012)\citenamefont {Helt},
  \citenamefont {Liscidini},\ and\ \citenamefont {Sipe}}]{helt2012does}%
  \BibitemOpen
  \bibfield  {author} {\bibinfo {author} {\bibfnamefont {L.~G.}\ \bibnamefont
  {Helt}}, \bibinfo {author} {\bibfnamefont {M.}~\bibnamefont {Liscidini}},\
  and\ \bibinfo {author} {\bibfnamefont {J.~E.}\ \bibnamefont {Sipe}},\
  }\bibfield  {title} {\bibinfo {title} {How does it scale? comparing quantum
  and classical nonlinear optical processes in integrated devices},\
  }\href@noop {} {\bibfield  {journal} {\bibinfo  {journal} {JOSA B}\ }\textbf
  {\bibinfo {volume} {29}},\ \bibinfo {pages} {2199} (\bibinfo {year}
  {2012})}\BibitemShut {NoStop}%
\bibitem [{\citenamefont {Liscidini}\ \emph {et~al.}(2012)\citenamefont
  {Liscidini}, \citenamefont {Helt},\ and\ \citenamefont
  {Sipe}}]{liscidini2012asymptotic}%
  \BibitemOpen
  \bibfield  {author} {\bibinfo {author} {\bibfnamefont {M.}~\bibnamefont
  {Liscidini}}, \bibinfo {author} {\bibfnamefont {L.}~\bibnamefont {Helt}},\
  and\ \bibinfo {author} {\bibfnamefont {J.}~\bibnamefont {Sipe}},\ }\bibfield
  {title} {\bibinfo {title} {Asymptotic fields for a hamiltonian treatment of
  nonlinear electromagnetic phenomena},\ }\href@noop {} {\bibfield  {journal}
  {\bibinfo  {journal} {Physical Review A}\ }\textbf {\bibinfo {volume} {85}},\
  \bibinfo {pages} {013833} (\bibinfo {year} {2012})}\BibitemShut {NoStop}%
\bibitem [{\citenamefont {Sipe}\ \emph {et~al.}(2004)\citenamefont {Sipe},
  \citenamefont {Bhat}, \citenamefont {Chak},\ and\ \citenamefont
  {Pereira}}]{sipe2004effective}%
  \BibitemOpen
  \bibfield  {author} {\bibinfo {author} {\bibfnamefont {J.}~\bibnamefont
  {Sipe}}, \bibinfo {author} {\bibfnamefont {N.~A.}\ \bibnamefont {Bhat}},
  \bibinfo {author} {\bibfnamefont {P.}~\bibnamefont {Chak}},\ and\ \bibinfo
  {author} {\bibfnamefont {S.}~\bibnamefont {Pereira}},\ }\bibfield  {title}
  {\bibinfo {title} {Effective field theory for the nonlinear optical
  properties of photonic crystals},\ }\href@noop {} {\bibfield  {journal}
  {\bibinfo  {journal} {Physical Review E}\ }\textbf {\bibinfo {volume} {69}},\
  \bibinfo {pages} {016604} (\bibinfo {year} {2004})}\BibitemShut {NoStop}%
\end{thebibliography}%

\end{document}